\documentclass[floatfix,twocolumn,showpacs,preprintnumbers,amsmath,amssymb,pra,superscriptaddress,longbibliography]{revtex4-1}
\usepackage{color}
\usepackage[usenames,dvipsnames,svgnames,table]{xcolor}
\usepackage[colorlinks=true,linkcolor=blue,urlcolor=blue,citecolor=blue]{hyperref}
\usepackage{mathtools}
\usepackage{graphicx}
\usepackage{dcolumn}
\usepackage{array}
\usepackage{lipsum}
\usepackage{bm}
\usepackage{subfigure}
\usepackage{amssymb}
\usepackage{multirow}
\usepackage{tabularx}
\usepackage{amsmath}
\usepackage{braket}
\graphicspath{{plots/}}
 \usepackage{lipsum}
\usepackage{mathrsfs}

\newcommand{\beq}{\begin{equation}}
\newcommand{\eeq}{\end{equation}}
\newcommand{\bea}{\begin{eqnarray}}
\newcommand{\eea}{\end{eqnarray}}

\begin{document}
%\underline{Draft version:}
\title{
Effective electronic forces and potentials\\ from \emph{ab initio} path integral Monte Carlo simulations 
}

\author{Tobias Dornheim}
\email{t.dornheim@hzdr.de}

\affiliation{Center for Advanced Systems Understanding (CASUS), D-02826 G\"orlitz, Germany}
\affiliation{Helmholtz-Zentrum Dresden-Rossendorf (HZDR), D-01328 Dresden, Germany}

\author{Panagiotis Tolias}
\affiliation{Space and Plasma Physics, Royal Institute of Technology (KTH), Stockholm, SE-100 44, Sweden}

\author{Zhandos A.~Moldabekov}

\affiliation{Center for Advanced Systems Understanding (CASUS), D-02826 G\"orlitz, Germany}
\affiliation{Helmholtz-Zentrum Dresden-Rossendorf (HZDR), D-01328 Dresden, Germany}

\author{Attila Cangi}

\affiliation{Center for Advanced Systems Understanding (CASUS), D-02826 G\"orlitz, Germany}
\affiliation{Helmholtz-Zentrum Dresden-Rossendorf (HZDR), D-01328 Dresden, Germany}

\author{Jan Vorberger}

\affiliation{Helmholtz-Zentrum Dresden-Rossendorf (HZDR), D-01328 Dresden, Germany}

\begin{abstract}
The rigorous description of correlated quantum many-body systems constitutes one of the most challenging tasks in contemporary physics and related disciplines. In this context, a particularly useful tool is the concept of effective pair potentials that take into account the effects of the complex many-body medium consistently. In this work, we present extensive, highly accurate \emph{ab initio} path integral Monte Carlo (PIMC) results for the effective interaction and the effective force between two electrons in the presence of the uniform electron gas (UEG). This gives us a direct insight into finite-size effects, thereby opening up the possibility for novel domain decompositions and methodological advances. In addition, we present unassailable numerical proof for an effective attraction between two electrons under moderate coupling conditions, without the mediation of an underlying ionic structure. Finally, we compare our exact PIMC results to effective potentials from linear-response theory, and we demonstrate their usefulness for the description of the dynamic structure factor. All PIMC results are made freely available online and can be used as a thorough benchmark for new developments and approximations.
\end{abstract}

\maketitle

\section{Introduction\label{sec:introduction}}

The accurate description of interacting quantum many-body systems~\cite{mahan1990many} constitutes one of the most active frontiers in a number of disciplines, such as theoretical physics, quantum chemistry, and material science.  The first groundbreaking insights into their collective behaviour have been obtained on the basis of well established theoretical approximations, such as the model of the \emph{dilute Bose gas}~\cite{RevModPhys.76.599} and the weakly coupled uniform electron gas (UEG) that can be described within the \emph{random phase approximation} (RPA)~\cite{pines2018theory}. Naturally, the situation becomes incomparably more complicated and interesting in the regime of moderate to strong coupling strength, where the aforementioned approximations break down. These theoretical challenges have facilitated the development of sophisticated numerical methods that are capable of capturing the rich interplay of quantum effects such as diffraction and quantum statistics with correlation effects as well as thermal excitations. A case in point is the \emph{ab initio} path integral Monte Carlo (PIMC) method~\cite{cep,Berne_JCP_1982,Takahashi_Imada_PIMC_1984}, which is a finite-temperature implementation of the quantum Monte Carlo (QMC) paradigm~\cite{anderson2007quantum}. Specifically, the PIMC approach has been of pivotal importance for our understanding of Bose liquids such as ultracold $^4$He~\cite{PhysRev.171.128,PhysRev.143.58,cep}. In fact, state-of-the-art implementations~\cite{boninsegni1,boninsegni2} allow for the quasi-exact simulation of up to $N\sim10^3-10^4$ particles, thereby giving us microscopic insights into important physical effects such as superfluidity~\cite{PhysRevB.36.8343,cep,PhysRevB.91.054503} and Bose-Einstein-condensation~\cite{griffin1993excitations,doi:10.7566/JPSJ.85.053001,PhysRevA.70.053614}.

Unfortunately, QMC simulations of quantum degenerate Fermi systems, such as the UEG~\cite{review} or ultracold Fermi atoms like $^3$He~\cite{Dornheim_SciRep_2022} are afflicted with the notorious fermion sign problem~\cite{troyer,dornheim_sign_problem}, which leads to an exponential increase in compute time with increasing the system size or decreasing the temperature. Therefore, QMC simulations of many-electron systems are usually restricted to rather small systems (typically $N\sim\mathcal{O}\left(10^1-10^2\right)$) even using state-of-the-art algorithms on modern high-performance computing clusters. At the same time, we note that these simulations are often remarkably accurate with as few as $N=14$ electrons~\cite{Chiesa_PRL_2006,dornheim_ML,dornheim_cpp,dornheim_prl,review,Dornheim_JCP_2021} and, in combination with dielectric methods (like the RPA) that become exact in the long-range limit, provide an adequate description over all length scales. This is particularly remarkable in the case of electrons, as the long-range Coulomb interaction is known to cause difficulties and divergencies in a number of approaches~\cite{quantum_theory,bonitz_book}.

In this context, a particularly important concept is the notion of \emph{effective pair interactions and forces}, which constitute a true mile stone in the description of quantum many-body systems. In general, we distinguish three distinct situations~\cite{quantum_theory,Kukkonen_Wilkins_PRB_1979}:
\begin{enumerate}
    \item The effective interaction between two test charges in a medium. The most prominent example is given by effective ion--ion interactions that are screened due to the presence of a medium consisting of free, nonideal electrons~\cite{pines,ceperley_potential,Ramazanov_PRE_2015,zhandos1, cpp_pot_22,cpp_pot_17}.
    \item The effective interaction between a test change and the surrounding medium, such as an ion embedded in a uniform electron gas~\cite{Bonev, PhysRevE.103.063206}.
    \item The effective interaction between two particles that are part of the surrounding medium. For example, the effective force $F_\textnormal{eff}(r)$ between two electrons in the UEG~\cite{Kukkonen_PRB_1979,Kukkonen_PRB_2021}, or the effective interaction between two bare nuclei in the interior of dense astrophysical objects~\cite{Alastuey_Jancovici_APJ_1978,doi:10.1063/1.873221}.
\end{enumerate}

In the present work, we present a detailed investigation of the third point based on \emph{ab initio} PIMC simulations of the UEG without any restrictions on the nodal structure of the thermal density matrix, and without any limitations based on linear-response theory. This gives us a direct insight into the remarkable absence of finite-size errors in QMC calculations of wavenumber resolved properties such as the static structure factor $S(q)$, and opens up new possibilities for future decompositions related to Kohn's celebrated principle of nearsightedness~\cite{Kohn_PNAS_2005}. In addition, we present unassailable numerical proof for an \emph{effective electronic attraction} between two electrons at certain parameters, without the mediation by an underlying ionic structure. Moreover, we investigate the impact and physical origin of different nonlinear effects~\cite{Dornheim_PRL_2020}, and demonstrate the hands-on utility of our new results for the interpretation of scattering experiments, and the description of the dynamic structure factor.

The paper is organized as follows: In Sec.~\ref{sec:theory}, we provide the required theoretical background, including a brief introduction to the UEG model and its dimensionless characteristic parameters (\ref{sec:UEG}), our implementation of the PIMC method (\ref{sec:PIMC}), the histogram estimation of the effective force and numerical computation of the corresponding effective pair interaction potential (\ref{sec:medium}), the quantum effective potential model of Kukkonen and Overhauser based on linear-response theory~\cite{Kukkonen_PRB_1979} (\ref{sec:LRT}) and the classical potential of mean force model based on distribution function theories (\ref{sec:classical}). Sec.~\ref{sec:results} is devoted to the in-depth discussion of our new PIMC simulation results, starting with a hands-on discussion of possible finite-size effects (\ref{sec:FSC}), proceeding with the investigation of the dependence of the effective interaction on thermodynamic parameters such as the density and the temperature (\ref{sec:dependence}), and culminating in the comparison of our PIMC data to different theoretical models~(\ref{sec:comparison}). Moreover, in Sec.~\ref{sec:applications}, we use our new PIMC results to estimate the effective potential (\ref{sec:potential}), and apply it to the interpretation of the previously reported \emph{roton feature} in the dynamic structure factor of the UEG~\cite{dornheim_dynamic} within the framework of the recently proposed \emph{electronic pair alignment} model~\cite{Dornheim_Nature_2022} (\ref{sec:spectrum}). We conclude our work with a brief summary of our main findings and a list of possible future investigations in Sec.~\ref{sec:summary}. Finally, an exact effective electronic potential connection between quantum linear response theory and classical distribution function theory is established in the appendix.

\section{Theory\label{sec:theory}}

\subsection{Uniform electron gas and effective system parameters\label{sec:UEG}}

Throughout this work, we consider an unpolarized electron gas (\emph{i.e.}, $N^\uparrow=N^\downarrow = N/2$, with $N$ being the total number of electrons) in a homogeneous rigid neutralizing background~\cite{loos,review,quantum_theory}. The Hamiltonian can then be expressed in the general form (we use Hartree atomic units throughout the main text)
\begin{eqnarray}\label{eq:Hamiltonian}
\hat{H} = -\frac{1}{2} \sum_{l=1}^N \nabla_l^2 + \frac{1}{2} \sum_{l\neq k}^N \phi_\textnormal{E}(\mathbf{\hat{r}}_l,\mathbf{\hat{r}}_k)\ ,
\end{eqnarray}
where $\phi_\textnormal{E}(\mathbf{r}_a,\mathbf{r}_b)$ is the usual Ewald pair interaction with $\mathbf{r}_a$ denoting the electron positions as discussed e.g.~in Ref.~\cite{Fraser_PRB_1996}. We note that we do not apply any additional external potential as it has been done in a number of recent works~\cite{Dornheim_PRL_2020,Dornheim_PRR_2021,Moldabekov_JCP_2021,Moldabekov_SciPost_2022}.

From a physical perspective, the UEG is fully characterized by two effective dimensionless parameters~\cite{Ott2018}: a) the density parameter (also known as Wigner-Seitz radius in the literature) $r_s=\overline{a}/a_\textnormal{B}$, with $\overline{a}$ and $a_\textnormal{B}$ being the mean distance to the nearest neighbour and the first Bohr radius, respectively, and b) the degeneracy temperature $\theta=k_\textnormal{B}T/E_\textnormal{F}$, with $E_\textnormal{F}$ denoting the Fermi energy~\cite{quantum_theory}, $T$ the temperature, and $k_\textnormal{B}$ the Boltzmann constant. It is common practice to introduce a coupling parameter $\Gamma = \braket{V}/\braket{K}$. The interaction energy and kinetic energy scale as $V\sim 1/r_s$ and $K\sim1/r_s^2$ in the degenerate case, respectively. Hence, it holds $\Gamma\sim r_s$, which means that the density parameter plays the role of a \emph{quantum coupling parameter} in the UEG. The range of metallic densities $r_s = 1,\dots,5$ can, thus, be characterized as \emph{moderately coupled} such that weak-coupling expansions like the random phase approximation (RPA)~\cite{quantum_theory} are not strictly applicable. For $r_s\to 0$, the UEG converges towards an ideal (i.e., noninteracting) Fermi gas. Conversely, for increasing  $r_s$ it forms an electron liquid~\cite{dornheim_electron_liquid,dornheim_dynamic} and, in the limit $r_s\gg1$, eventually a Wigner crystal~\cite{PhysRevB.69.085116,Azadi_Wigner_2022} .

In addition, the degeneracy temperature $\theta$ constitutes a straightforward measure for the importance of quantum degeneracy effects, with $\theta\gg 1$ and $\theta\ll 1$ corresponding to the classical limit~\cite{dornheim2022wolf} and fully degenerate case, respectively. In the present work, we are primarily interested in the intermediate regime of $r_s\sim\theta\sim 1$. These so-called \emph{warm dense matter} (WDM) conditions are characterized by the complicated interplay of thermal, Coulomb coupling, and quantum degeneracy effects and naturally occur in a number of astrophysical applications such as giant planet interiors~\cite{Benuzzi_Mounaix_2014,militzer1}. Moreover, these extreme states can be realized in experiments at large-scale research facilities using different techniques, see e.g.~the topical review by Falk~\cite{falk_wdm}.

For completeness, we note that a third relevant parameter is the spin-polarization $\xi=(N^\uparrow-N^\downarrow)/N$, where $\xi=0$ and $\xi=1$ correspond to the unpolarized (paramagnetic) and fully polarized (ferromagnetic) limits. In the present work, we restrict ourselves to the case of $\xi=0$; a detailed investigation of the dependence of the effective force and interaction on the spin-polarization is particularly relevant in the presence of an external magnetic field~\cite{Dornheim_PRE_2021} and constitutes an important project for future works.

\subsection{Path integral Monte Carlo\label{sec:PIMC}}

The rigorous theoretical description of the UEG in the WDM regime, the primary focus of the present work, constitutes a formidable challenge, as a number of complex effects~\cite{wdm_book,new_POP,review} need to be taken into account. In this context, the most promising method is the quantum Monte Carlo (QMC) technique~\cite{Foulkes_RMP_2001}, as it can, in principle, give an exact solution to the full quantum many-body problem of interest without any empirical input (such as the \textit{a priori} unknown exchange--correlation functional of density functional theory~\cite{Burke_JCP_2012}). At finite temperature, the most widely used QMC method is the path integral Monte Carlo (PIMC) approach~\cite{cep,Berne_JCP_1982,Takahashi_Imada_PIMC_1984}, which is based on the exact isomorphism between the quantum system and a classical system of interacting ring-polymers~\cite{Chandler_Wolynes_JCP_1981}.
A detailed introduction to PIMC has already been presented elsewhere~\cite{cep,boninsegni1}. We, therefore, restrict the presentation here to a concise discussion of the underlying idea.

Let us consider the partition function of $N$ unpolarized electrons in the canonical ensemble (i.e., the inverse temperature $\beta=1/k_\textnormal{B}T$, number density $n=N/\Omega$ and volume $\Omega$ are fixed), which can be expressed in coordinate space as
\begin{widetext}
\begin{eqnarray}\label{eq:Z}
Z_{\beta,N,\Omega} = \frac{1}{N^\uparrow! N^\downarrow!} \sum_{\sigma^\uparrow\in S_{N^\uparrow}} \sum_{\sigma^\downarrow\in S_{N^\downarrow}} \textnormal{sgn}(\sigma^\uparrow,\sigma^\downarrow)\int d\mathbf{R} \bra{\mathbf{R}} e^{-\beta\hat H} \ket{\hat{\pi}_{\sigma^\uparrow}\hat{\pi}_{\sigma^\downarrow}\mathbf{R}}\ .
\end{eqnarray}
\end{widetext}
Here, $\mathbf{R}=(\mathbf{r}_1,\dots ,\mathbf{r}_N)^T$ contains the coordinates of both spin-up and spin-down electrons. The sums run over all possible permutations $\sigma^i$ from the respective permutation group $S_{N^i}$ ($i\in\{\uparrow,\downarrow\}$), where $\hat{\pi}_{\sigma^i}$ is the corresponding permutation operator. This ensures the correct fermionic anti-symmetry under the exchange of particle coordinates. The basic idea behind the PIMC method is to perform a Trotter decomposition~\cite{trotter} of the \textit{a priori} unknown matrix elements of the density operator $\hat{\rho}=e^{-\beta\hat{H}}$. This leads to a product of $P$ density matrices, which have to be evaluated at $P$-times the original temperature. If this parameter is chosen sufficiently large, one can evaluate the density matrix within a suitable high-temperature approximation. In the present work, we employ the simple primitive factorization
\begin{eqnarray}\label{eq:primitive}
e^{-\epsilon\hat{H}} \approx e^{-\epsilon\hat{K}}e^{-\epsilon\hat{V}}\ ,
\end{eqnarray}
with $\epsilon=\beta/P$, which becomes exact in the large-$P$ limit as $\mathcal{O}\left(P^{-2}\right)$~\cite{brualla_JCP_2004}. The convergence with $P$ has been carefully checked (we find $P=200$ sufficient at the present conditions), and our results are, therefore, \emph{quasi-exact}. For completeness, we note that higher-order factorizations of $\hat{\rho}$ have been studied in the literature~\cite{sakkos_JCP_2009,ZILLICH2015111} and become important for PIMC simulations in the low-temperature regime.

\begin{figure}\centering
\includegraphics[width=0.475\textwidth]{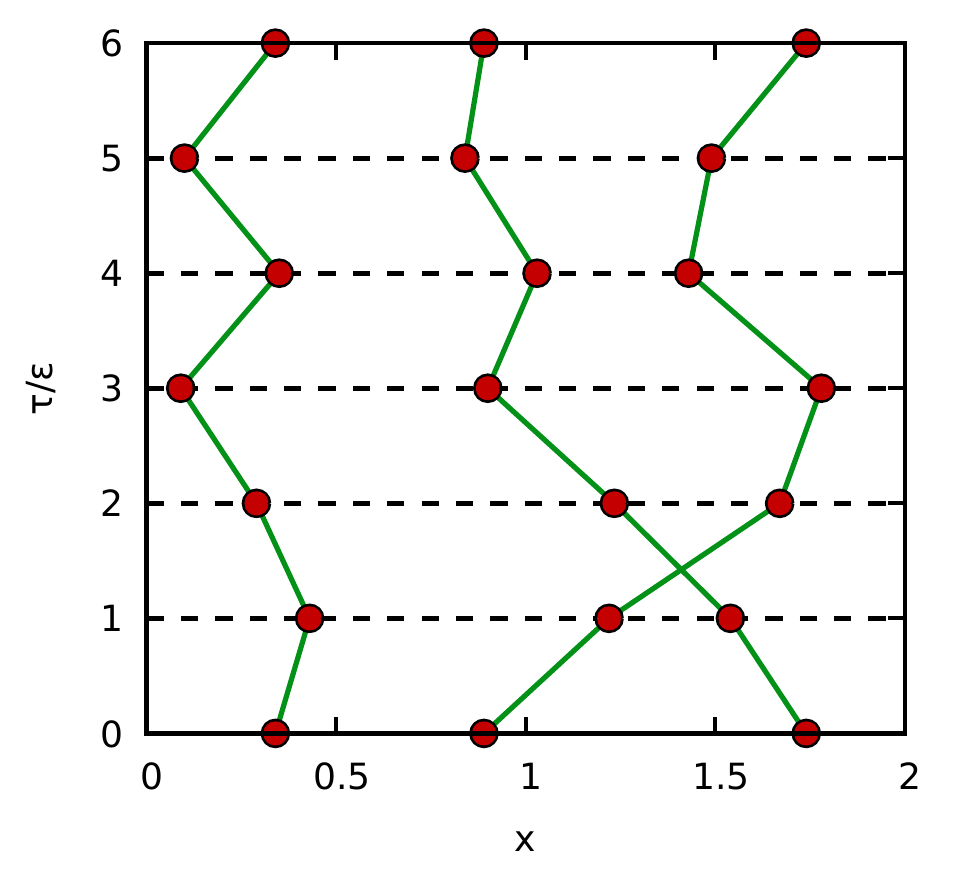}
\caption{\label{fig:PIMC}
Schematic illustration of the PIMC method. Shown is a configuration of $N=3$ particles in the $\tau$-$x$-plane. The horizontal dashed lines correspond to the $P$ time slices with the imaginary-time step $\epsilon=\beta/P$. Each particle is represented by a \emph{closed path} along the $\tau$-direction. The two particles on the right are involved in a pair-exchange cycle~\cite{Dornheim_permutation_cycles}, and the corresponding weight is thus negative, $W(\mathbf{X})<0$.
}
\end{figure} 

The PIMC representation of the partition function can then be expressed in the compact form
\begin{eqnarray}\label{eq:Z_PIMC}
Z = \int \textnormal{d}\mathbf{X}\ W(\mathbf{X})\ ,
\end{eqnarray}
where the new meta-variable $\mathbf{X}=(\mathbf{R}_0,\dots ,\mathbf{R}_{P-1})^T$ includes $P$ sets of coordinates of all $N$ particles, and the integration includes the sum over all permutations. A graphical illustration of Eq.~(\ref{eq:Z_PIMC}) is given in Fig.~\ref{fig:PIMC}, where we show a configuration of $N=3$ particles. Evidently, each particle is now represented by an entire \emph{path} along the imaginary time $\tau\in[0,\beta]$; this is the origin of the ring-polymers from the celebrated classical isomorphism by Chandler and Wolynes~\cite{Chandler_Wolynes_JCP_1981}. The basic idea of the PIMC method is to use a variation of the Metropolis algorithm~\cite{metropolis} to generate a Markov chain of random configurations $\{\mathbf{X}\}$ that are distributed according to the probability $P(\mathbf{X})=W(\mathbf{X})/Z$, with $W(\mathbf{X})$ being the corresponding configuration weight. For bosons (such as ultracold atoms like $^4$He~\cite{cep}) or hypothetical distinguishable quantum particles (often denoted as \emph{boltzmannons}), this procedure is straightforward. Indeed, using modern Monte Carlo sampling techniques~\cite{boninsegni1,boninsegni2} quasi-exact results for up to $N\sim10^4$ particles for such system are obtained. 

In stark contrast, the situations is incomparably more complicated in the case of fermions (such as the electrons in which we are interested in the present work, but also ultracold atoms like $^3$He~\cite{Dornheim_SciRep_2022}). Here, the configuration weight $W(\mathbf{X})$ is \emph{negative} for an odd number of pair exchanges in a given PIMC configuration, c.f.~the sign function $\textnormal{sgn}(\sigma^\uparrow,\sigma^\downarrow)$ in Eq.~(\ref{eq:Z}). A corresponding example is shown in Fig.~\ref{fig:PIMC}, where the two particles on the right are involved in a so-called \emph{permutation cycle}~~\cite{Dornheim_permutation_cycles}, i.e., a trajectory with more than a single particle in it.

From a practical perspective, the fermionic antisymmetry under the exchange of particle coordinates directly implies that $P(\mathbf{X})$ cannot be interpreted as a probability in the Monte Carlo procedure. To overcome this obstacle, we define a modified configuration space as
\begin{eqnarray}\label{eq:modified}
P'(\mathbf{X}) &=& \frac{|W(\mathbf{X})|}{Z'} \\ \nonumber
Z' &=& \int \textnormal{d}\mathbf{X}\ |W(\mathbf{X})|\ ,
\end{eqnarray}
and randomly generate the paths according to $P'(\mathbf{X})$.
In particular, Eq.~(\ref{eq:modified}) means we carry out a \emph{bosonic} PIMC simulation; the exact fermionic expectation value of interest is subsequently extracted by evaluating the ratio
\begin{eqnarray}\label{eq:ratio}
\braket{\hat{A}} = \frac{\braket{\hat{A}\hat{S}}'}{\braket{\hat{S}}'}\ .
\end{eqnarray}
Here $S(\mathbf{X})=W(\mathbf{X})/|W(\mathbf{X})|$ is the sign of a particular configuration, and the denominator in Eq.~(\ref{eq:ratio}) is commonly being referred to as the \emph{average sign}.
The sign changes in both the enumerator and the denominator in Eq.~(\ref{eq:ratio}) lead to a cancellation of positive and negative terms, which is the root cause of the notorious \emph{fermion sign problem}~\cite{troyer,dornheim_sign_problem,Dornheim_JPA_2021}. In practice, the latter manifests as an exponential increase in the required compute time with important system parameters such as the system size $N$ or the inverse temperature $\beta$~\cite{dornheim_sign_problem}.

Due to its fundamental nature, a number of strategies have been proposed to tackle the sign problem. Ceperley and others~\cite{Ceperley1991,MILITZER201913,Militzer_PRL_2015,Brown_PRL_2013} have introduced the \emph{fixed-node approximation} to the PIMC method, sometimes referred to as \emph{restricted} PIMC (RPIMC) in the literature. On the one hand, RPIMC formally avoids the exponential bottleneck and, therefore, allows for simulations at parameters that are unfeasible otherwise. Yet, this advantage comes at the cost of an uncontrolled, systematic approximation, effectively breaking the \emph{quasi-exact} nature of PIMC. Indeed, Schoof \emph{et al.}~\cite{Schoof_PRL_2015} have been able to assess the accuracy of RPIMC at WDM conditions based on exact configuration PIMC (CPIMC) simulations. They have found systematic deviations in the exchange--correlation energy between the two methods in excess of $10\%$. This surprisingly large bias has subsequently been substantiated in independent studies~\cite{Malone_PRL_2016,Joonho_JCP_2021,review,dornheim_POP}.

This unsatisfactory situation has sparked a recent surge of activity regarding the development of fermionic QMC methods at finite temperature~\cite{Blunt_PRB_2014,Malone_PRL_2016,Rubenstein_auxiliary_finite_T,Malone_JCP_2015,Dornheim_NJP_2015,dornheim_jcp,Dornheim_JCP_2020,Yilmaz_JCP_2020,Joonho_JCP_2021,Filinov_CPP_2021,universe8020079}; see Ref.~\cite{dornheim_POP} for a topical discussion of some of these developments. In the present work, we use the standard PIMC method that is heavily afflicted by the sign problem in the WDM regime~\cite{dornheim_sign_problem}. As a consequence, our simulations are computationally expensive (we use up to $\sim10^5$CPUh for a single run, distributed among $\sim10^4$ cores on a modern high-performance computing cluster), but \emph{exact} within the stated Monte Carlo error bars. All PIMC data of our study are freely available online~\cite{repo} and can serve as a rigorous benchmark for the development of new methods and approximations.

\subsection{Effective electronic force in an electronic medium\label{sec:medium}}

\begin{figure}\centering
\includegraphics[width=0.475\textwidth]{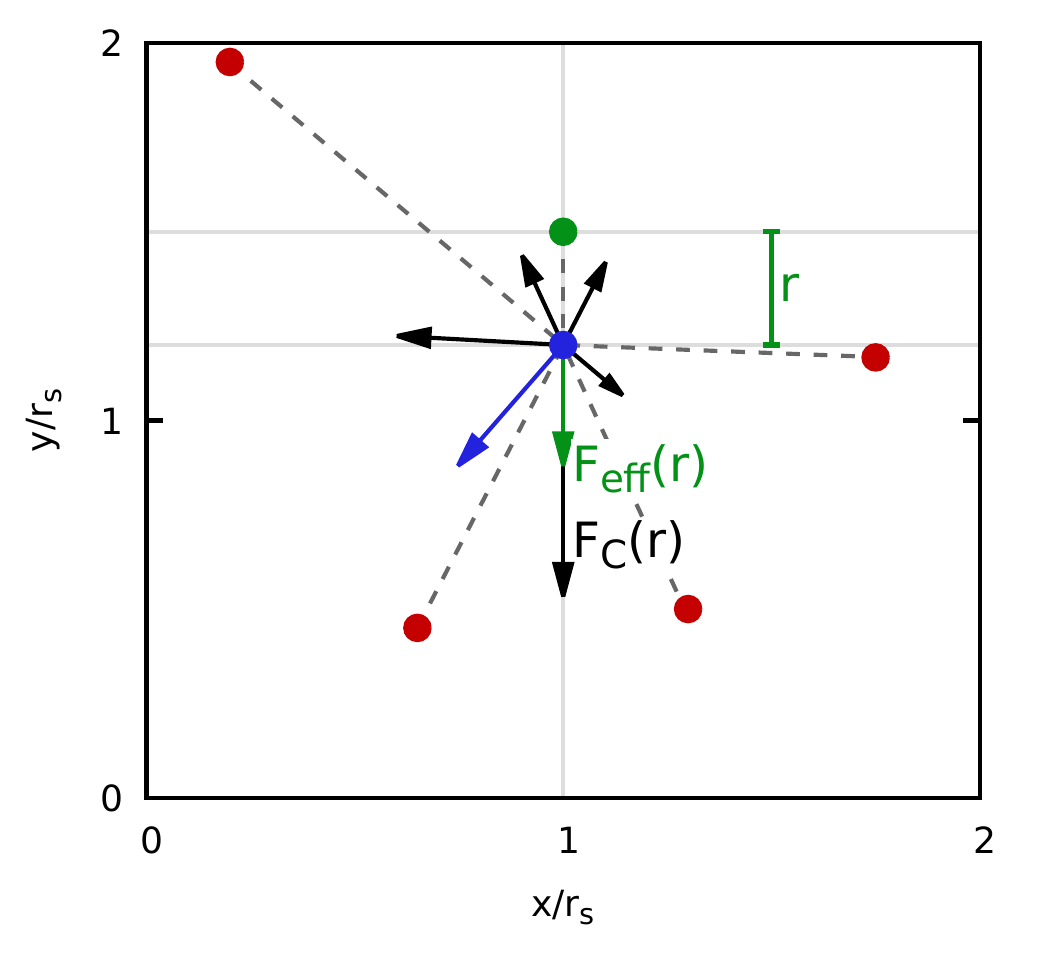}
\caption{\label{fig:rectangle}
Schematic illustration of the effective force $F_\textnormal{eff}(r)$. The black arrows indicate the bare Coulomb forces, and the blue arrow shows the total pair force on the blue reference particle. The green bead depicts a second particle at a distance $r$, and the green arrow the corresponding effective force $F_\textnormal{eff}(r)$, i.e., the projection of the total force along the direction of the difference vector $\mathbf{r}$. 
}
\end{figure} 

The central goal of the present work is to accurately compute the \emph{effective force} $F_\textnormal{eff}(r)$ between two electrons immersed in a UEG, i.e., the average force between two particles at a certain distance $r$. By definition, this quantity must not include the direct exchange and correlation effects between the two electrons themselves~\cite{quantum_theory}.
This is schematically illustrated in Fig.~\ref{fig:rectangle}, where we show a configuration of $N=6$ particles in coordinate space. The blue bead represents an (arbitrarily chosen) \emph{reference particle} at a position $\mathbf{r}_a$. The black arrows correspond to the individual pair forces on that particle
\begin{eqnarray}
\mathbf{F}_{a,l} = -\nabla_a \phi_\textnormal{E}(\mathbf{r}_a,\mathbf{r}_l)\ ,
\end{eqnarray}
computed from the gradient of the Ewald pair potential  $\phi_\textnormal{E}(\mathbf{r}_a,\mathbf{r}_l)$ ~\cite{Fraser_PRB_1996,dornheim_jcp} mentioned above.
The total force on the reference particle is depicted as the blue arrow in Fig.~\ref{fig:rectangle}, and is given by
\begin{eqnarray}\label{eq:F}
\mathbf{F}_{a} = \sum_{l\neq a}^N \mathbf{F}_{a,l}\ .
\end{eqnarray}
To compute the sought-after \emph{effective force} between a second particle (green bead in Fig.~\ref{fig:rectangle}) at the position $\mathbf{r}_b$, with $\mathbf{r}\equiv\mathbf{r}_a-\mathbf{r}_b$ and $r\equiv|\mathbf{r}|$, we project the total force $\mathbf{F}_a$ onto the direction of $\mathbf{r}$,
\begin{eqnarray}
{F}_\textnormal{eff}(r) = \frac{1}{r} \mathbf{r}\cdot\mathbf{F}_a\ ,
\end{eqnarray}
see the green arrow. The relatively short distance between the green and blue particles leads to a comparably strong bare Coulomb force, i.e., the black arrow pointing downwards. The presence of the surrounding electronic medium effectively reduces this force. In other words, the two red particles at the bottom of the figure \emph{push} the blue reference particle towards the green particle and, thus, reduce the repulsion between them.

For bosons, and distinguishable Boltzmann particles,
the estimation of this effective force in our PIMC simulations is then straightforward. On each imaginary-time slice, one has to i) compute the full force on each particle $\mathbf{F}_l$ and ii) project that force along the direction of all $N-1$ particle pairs. Note that step ii) is repeated for all $N$ particles on each slice. In practice, this results in a \emph{histogram estimator}, where one can measure $F_\textnormal{eff}(r)$ only for certain distances $r=r_{l,k}$ in each configuration.

For fermions, on the other hand, the situation is somewhat more subtle. Specifically, the simple histogram estimator does not take into account fermionic exchange effects, but only implicitly corrects for correlations in the bosonic reference system, cf.~Sec.~\ref{sec:PIMC} above. A more rigorous expression for the effective force is given by
\begin{eqnarray}\label{eq:rigorous2}
F_\textnormal{eff}(r) = \frac{1}{\left<\sum_{l\neq k}^N \delta_{r_{lk},r} \right>} \left<
\sum_{l\neq k}^N \frac{\mathbf{r}_{lk}\cdot\mathbf{F}_{k}}{r_{lk}}\delta_{r_{lk},r}
\right>\ ,
\end{eqnarray}
where the second term corresponds to the actual measurement of the force for certain configurations, and the first term to the corresponding normalization. It is easy to see that this definition exactly fulfills the notion that $F_\textnormal{eff}(r)$ must exclude any correlation effects between the two reference particles.
We note that Eq.~(\ref{eq:rigorous2}) is the most general form and holds for every type of quantum statistics; the histogram estimator discussed above is recovered for bosons, Boltzmann particles, and in the classical limit. Re-calling the evaluation of a fermionic expectation value given in Eq.~(\ref{eq:ratio}) above,
Eq.~(\ref{eq:rigorous2}) becomes
\begin{eqnarray}\label{eq:rigorous}
F_\textnormal{eff}(r) = \frac{1}{\left<\hat{S}\sum_{l\neq k}^N \delta_{r_{lk},r} \right>'} \left<\hat{S}
\sum_{l\neq k}^N \frac{\mathbf{r}_{lk}\cdot\mathbf{F}_{k}}{r_{lk}}\delta_{r_{lk},r}
\right>'\ ;\quad\quad
\end{eqnarray}
we have already made use of the fact that the expectation values of the average sign cancel from both terms. Evidently, both the normalization and the actual force measurement in Eq.~(\ref{eq:rigorous}) sensitively depend on fermionic exchange effects via the sign operator $\hat{S}$ that has to be evaluated in each bosonic reference configuration. Specifically, this consistently fermionic definition of the normalization distinguishes Eq.~(\ref{eq:rigorous2}) from the simple histogram, which can be expressed as 
\begin{eqnarray}\label{eq:rigorous3}
F_\textnormal{hist}(r) = \frac{1}{\left<\sum_{l\neq k}^N \delta_{r_{lk},r} \right>'}\frac{ \left<\hat{S}
\sum_{l\neq k}^N \frac{\mathbf{r}_{lk}\cdot\mathbf{F}_{k}}{r_{lk}}\delta_{r_{lk},r}
\right>'}{\braket{\hat{S}}'}\ .\quad 
\end{eqnarray}
This has a profound impact on the force between two electrons of the same spin-orientation for $r\lesssim r_s$, as it is shown in Sec.~\ref{sec:comparison} below.

Finally, we note that results for $F_\textnormal{eff}(r)$ can be used to compute a corresponding interaction potential by numerically solving the one-dimensional integral
\begin{eqnarray}\label{eq:Phi}
\phi_\textnormal{eff}(r) = \int_{\infty}^r \textnormal{d}r'\ F_\textnormal{eff}(r')\ .
\end{eqnarray}

\subsection{Linear response theory and the Kukkonen-Overhauser potential\label{sec:LRT}}

Within linear response theory, the effective potential between any two electrons in the UEG was first derived by Kukkonen and Overhauser (KO)~\cite{Kukkonen_PRB_1979}, see also the more recent Refs.~\cite{Kukkonen_PRB_2021,quantum_theory}. Their formulation requires the knowledge of accurate dynamic or static local field corrections~\cite{Richardson_1994} and provides a convenient reference point for comparison of our PIMC results.

Aside from the uniform electron liquid, applications concern also the theory of superconductivity~\cite{Richardson_1997a,Richardson_1997b,Takada_PRB_1993}.

It should be first noted that the general KO treatment includes a contribution that describes the effects of lattice screening on the effective electronic potential; a term that is important for real metals (see the phonon mediated effective electronic attraction that drives a system to a superconducting BCS state), but is non-existent for the UEG~\cite{Kukkonen_PRB_1979,quantum_theory}. For the unpolarized case of interest, the spin-resolved Fourier transformed KO effective potential is given by a two-electron operator in spin space that reads as~\cite{quantum_theory}
\begin{align}\label{eq:KOgeneral}
\hat{\Phi}^{\mathrm{KO}}_{\mathrm{eff}}(q,\omega)&=\Phi(q)+\left\{\Phi(q)\left[1-G_+(q,\omega)\right]\right\}^2\chi_{\mathrm{nn}}(q,\omega)+\nonumber\\&\,\,\,\,\,\,\,\,\hat{\boldsymbol{\sigma}}_1\hat{\boldsymbol{\sigma}}_2\left\{\Phi(q)G_-(q,\omega)\right\}^2\chi_{\mathrm{ss}}(q,\omega)\,,
\end{align}
where $\chi_{\mathrm{nn}}(q,\omega)$ is the density-density response function, $\chi_{\mathrm{ss}}(q,\omega)$ is the longitudinal spin-spin response function, $G_+(q,\omega),G_-(q,\omega)$ are the spin symmetric and spin antisymmetric dynamic local field corrections (DLFC), $\hat{\boldsymbol{\sigma}}_1$, $\hat{\boldsymbol{\sigma}}_2$ are the relevant spin-operators and $\Phi(q)=4\pi/q^2$ is the Fourier transformed bare Coulomb interaction.

In the unpolarized case, the latter contribution cancels out when interest lies in the effective potential between any two electrons. Restricting oneself to the static limit, the KO effective potential ultimately reads as~\cite{quantum_theory}
\begin{eqnarray}\label{eq:KO}
\Phi^{\mathrm{KO}}_{\mathrm{eff}}({q})=\frac{4\pi}{q^2}+\left[\frac{4\pi}{q^2}\left(1-G({q})\right)\right]^2\chi({q}) \ ,
\end{eqnarray}
with $\chi_{\mathrm{nn}}(q,0)=\chi({q})$ the static density-density response that describes the linear density response of the full paramagnetic UEG to an external perturbation, \emph{i.e.}, of both spin-up and spin-down electrons, and where $G_+(q,0)=G(q)$ is the static local field correction (SLFC).

The static density-density response is conveniently expressed as~\cite{kugler1}
\begin{eqnarray}\label{eq:chi_tot}
\chi(q) = \frac{\chi_0(q)}{1-\frac{4\pi}{q^2}\left[1-G(q)\right]\chi_0(q)}\ ,
\end{eqnarray}
where $\chi_0(q)$ is the density response of a non-interacting Fermi gas at the same conditions that can be readily computed~\cite{quantum_theory}. Therefore, the effective KO potential is given exclusively as a functional of the SLFC $G(q)$, which contains the full wavenumber-resolved information about static exchange--correlation effects; setting $G(q)\equiv 0$ thus corresponds to the RPA. In the present work, we use the highly accurate neural-net representation of $G(q;r_s,\theta)$ by Dornheim \emph{et al.}~\cite{dornheim_ML} that is based on extensive PIMC simulation data. For completeness, we note that the recent analytical representation of $G(q;r_s,\theta)$~\cite{Dornheim_PRB_ESA_2021} within the effective static approximation~\cite{Dornheim_PRL_2020_ESA} would be equally suitable for this purpose.

Finally, it is evident that, within linear response theory, the effective force between two electrons in the UEG is given by
\begin{eqnarray}\label{eq:F_KO}
F_{\mathrm{eff}}^\textnormal{KO}(r) = -\frac{\textnormal{d}}{\textnormal{d}r}\phi_{\mathrm{eff}}^\textnormal{KO}(r)\ ,
\end{eqnarray}
which can be evaluated both using the SLFC and within the RPA.

\subsection{Classical theory of liquids and the potential of mean force}\label{sec:classical}

The effective potential between two particles is an important quantity in the classical theory of liquids, where it has been coined as the potential of mean force~\cite{liquid_state_Cole,liquid_state_Egelstaff,liquid_state_McQuarrie,liquid_state_Croxton}. The effective potential was formally introduced in a seminal article by Kirkwood~\cite{Kirkwood1935}, although its physical meaning had also been discussed in earlier works of Onsager, Einstein and Smoluchowski~\cite{Onsager1933}. In the canonical ensemble, it is elementary to show that $F_{\mathrm{cl}}(r)=\nabla_{\boldsymbol{r}}\ln{[g(r)]}$ which leads to~\cite{liquid_state_McQuarrie,liquid_state_Croxton,Kirkwood1935} 
\begin{eqnarray}\label{eq:phi_cl}
w^{(2)}(r)=\phi_\textnormal{cl}\left[g(r)\right]=-\frac{1}{\beta}\textnormal{ln}\left[g(r)\right]\,,
\end{eqnarray}
where $\beta{w}^{(2)}(r)$ is the interaction potential between two particles held apart at a fixed distance $r$ with the contributions of the remaining $N-2$ particles ensemble averaged over all canonical configurations. Note that the formal equation and its physical interpretation can be generalized to arbitrary correlation order, \emph{i.e.}, $\beta{w}^{(s)}(\boldsymbol{r^{s}})=\textnormal{ln}\left[g^{s}(\boldsymbol{r^{s}})\right]$, with $g^{(s)}(\boldsymbol{r^{s}})$ the $s-$correlation function and $\boldsymbol{r^{s}}=(\boldsymbol{r}_1,\boldsymbol{r}_2,...,\boldsymbol{r}_s)$ the reduced configuration vector~\cite{liquid_state_McQuarrie}.

The potential of mean force $\beta{w}^{(2)}(r)$ is invoked in the non-probabilistic interpretation of the Kirkwood superposition approximation which amounts to the assumption that the potential of mean force for a triplet of particles $\beta{w}^{(3)}(r)$ is pair-wise additive~\cite{liquid_state_Cole,liquid_state_McQuarrie}, the physical interpretation of the BGY hierarchy (the thermodynamic equilibrium version of the BBGKY hierarchy) which can be viewed as a force balance equation~\cite{liquid_state_Croxton} as well as the physical interpretation of the Kirkwood charging equation which can also be viewed as an expression of force balance~\cite{liquid_state_McQuarrie}. Furthermore, the short-distance determination of the potential of mean force from computer simulations is the most difficult step in the indirect extraction of bridge functions~\cite{Restrepo_1992,Tomazic_2014,LuccoCastello_2021}; the key diagrammatic object of the integral equation theory of classical liquids. It is worth noting that Eq.~(\ref{eq:phi_cl}) generalizes the Boltzmann relation $g(r)=\exp{\left[-\beta\phi(r)\right]}$, that is exact in the dilute classical limit $n\to0$, to any order in density\,\cite{liquid_state_Lee,liquid_state_Santos}.

Moreover, the potential of mean force leads to the definition of a screening potential $\beta{H}(r)=\beta\phi(r)-\beta{w}^{(2)}(r)$, which is a measure of the residual beyond-pair potential interaction that remains finite even at contact $r=0$. The screening potential is a very important quantity for the classical one-component plasma~\cite{LuccoCastello2022}, for which its short range behavior can be well-approximated with the aid of the Widom even-power expansion for the cavity distribution function $y(r)=g(r)\exp{[\beta{\phi(r)}]}$~\cite{Widom1963}, the Jancovici theoretical result for the second-order term~\cite{Jancovici1977} and the consideration of the interaction-site molecule as an extreme case of binary ionic mixture for the zero-order term~\cite{Rosenfeld1992}. In particular, the leading order term in the enhancement of nuclear reaction rates due to effective interactions in the interior of dense astrophysical objects can be approximated by $\exp{[\beta{H}(0)]}$\,\cite{Alastuey_Jancovici_APJ_1978,doi:10.1063/1.873221}, when the semi-classical Wentzel-Kramers-Brillouin (WKB) approximation is applicable for the solution of the relevant two-particle Schr\"odinger equation\,\cite{Salpeter1969,Ichimaru_Rev1993}.

Finally, the physical trends of the potential of mean force can be easily deduced from Eq.~(\ref{eq:phi_cl}) and the known functional behavior of the pair correlation function. At weak coupling, $g(r)$ exhibits a monotonic behavior that is characterized by a pure exponential asymptotic decay to unity. Therefore, the potential of mean force is monotonic and positive everywhere. At strong coupling, $g(r)$ has a well-defined extended correlation void and then exhibits a non-monotonic profile that is characterized by an exponentially damped oscillatory behavior around unity. As a consequence, the potential of mean force is highly non-monotonic and features alternating regions of effective attraction and repulsion that grow weaker with the distance. Identical to $g(r)$, the transition from monotonic to non-monotonic behavior formally defines a line in the phase diagram known as the Fisher-Widom~\cite{Evans1993} or the Kirkwood line~\cite{Carvalho1999}, depending on the mathematical origin of the transition. 

As deduced from Eq.~(\ref{eq:phi_cl}), the maximum of the first coordination shell corresponds exactly to the global maximum of the attractive effective force. Any particle essentially requires another particle to be quasi-localized therein for energy minimization purposes and the background ensures that another particle is attracted towards that region. This is followed by a local maximum of the repulsive effective force that corresponds exactly to the minimum of the first coordination shell. Essentially, the quasi-localization of a particle at the coordination shell maximum is primary responsible for this repulsive force that prevents particles from occupying the nearby coordination shell minimum. In this manner, the unfolding of short range order generates alternating effective attractions and repulsions.

In the present work, we use highly accurate \emph{ab initio} PIMC results for the UEG to assess the validity of these classical trends in a strongly coupled quantum system.

\section{Results\label{sec:results}}

All PIMC simulation results shown in this work have been obtained on the basis of the \emph{extended ensemble} scheme that has been introduced in Ref.~\cite{Dornheim_PRB_nk_2021}. A repository of our simulation data is freely available online~\cite{repo}.

\subsection{Finite-size effects\label{sec:FSC}}

\begin{figure*}\centering
\includegraphics[width=0.475\textwidth]{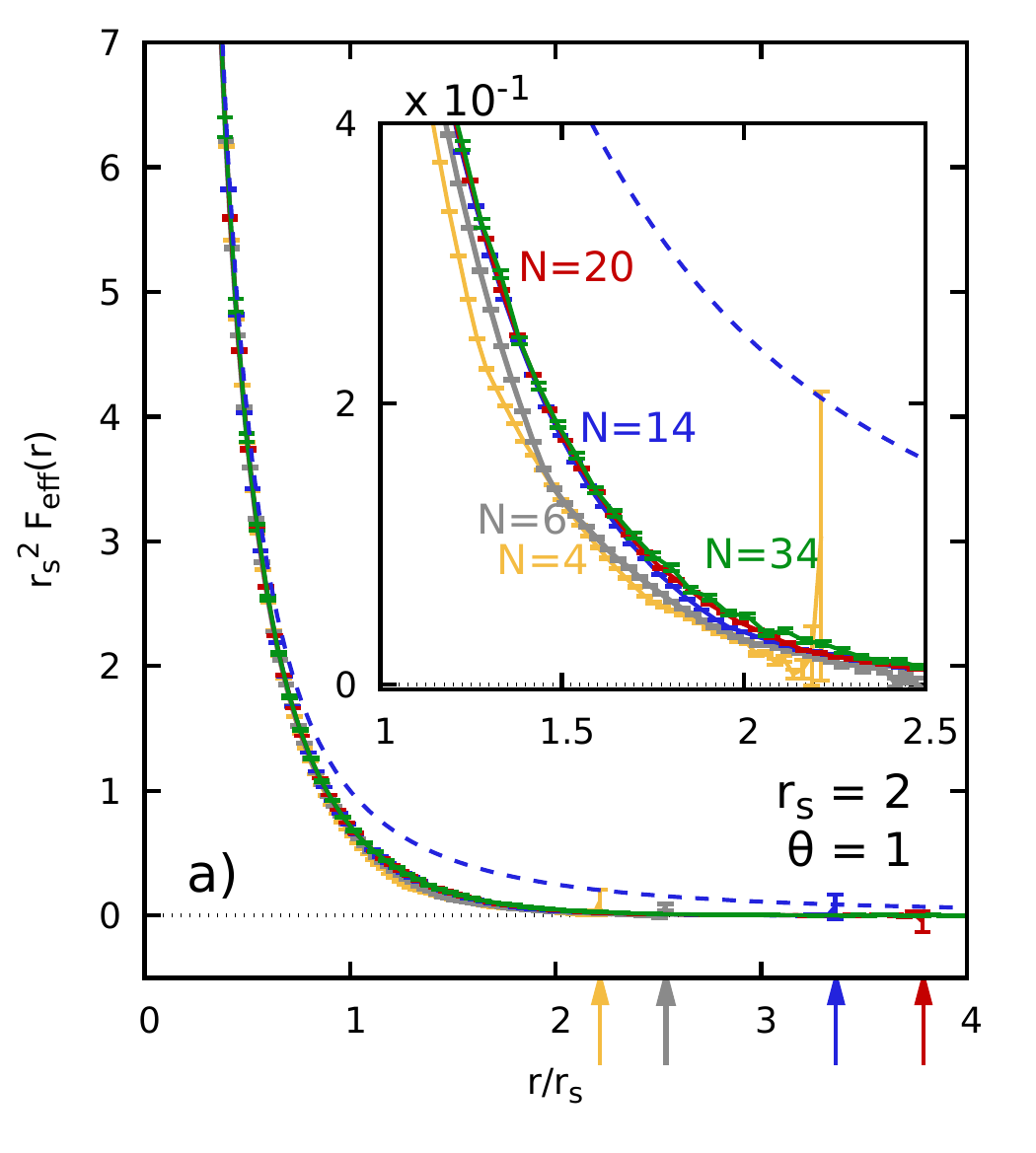}\includegraphics[width=0.475\textwidth]{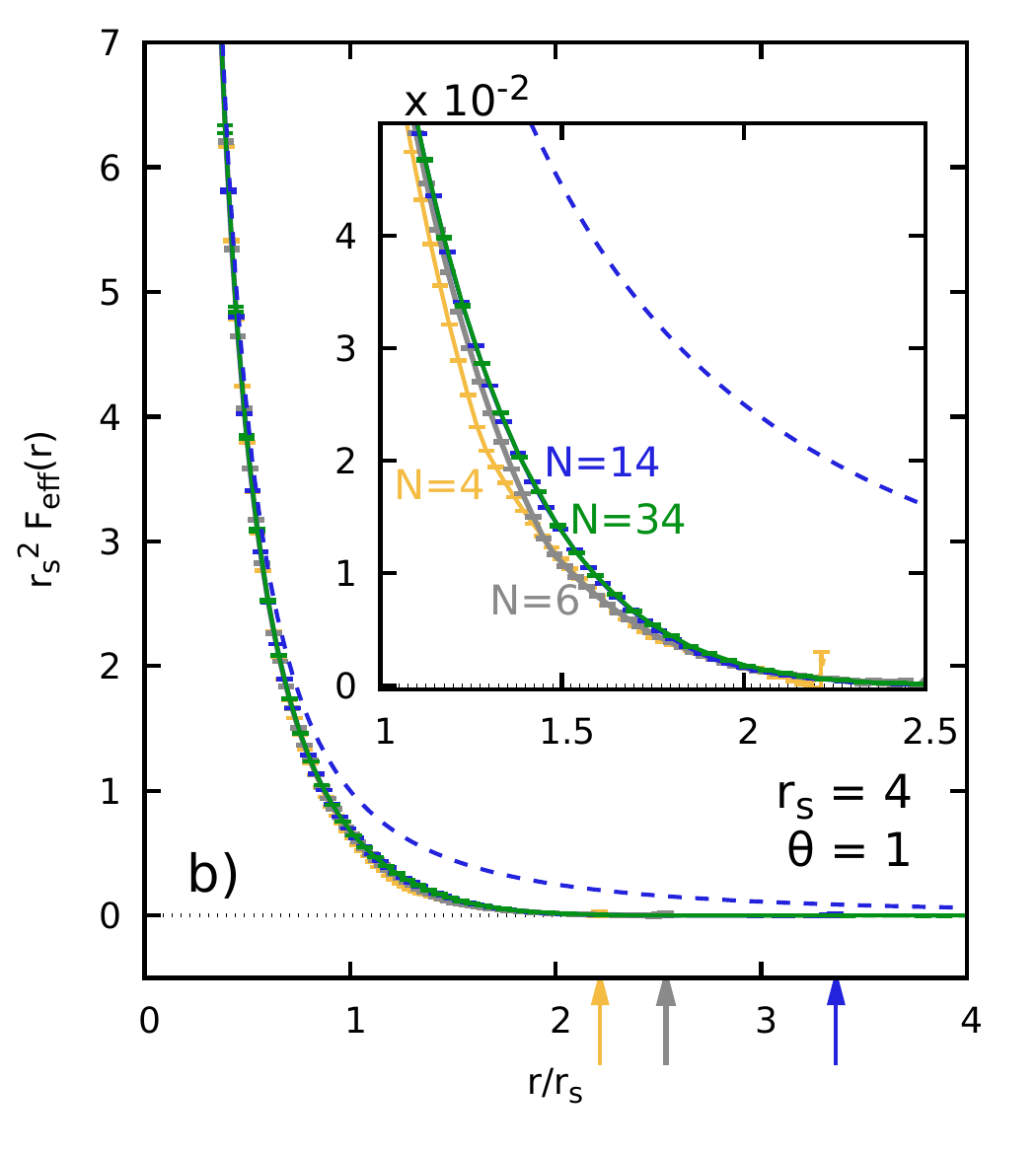}\\
\includegraphics[width=0.475\textwidth]{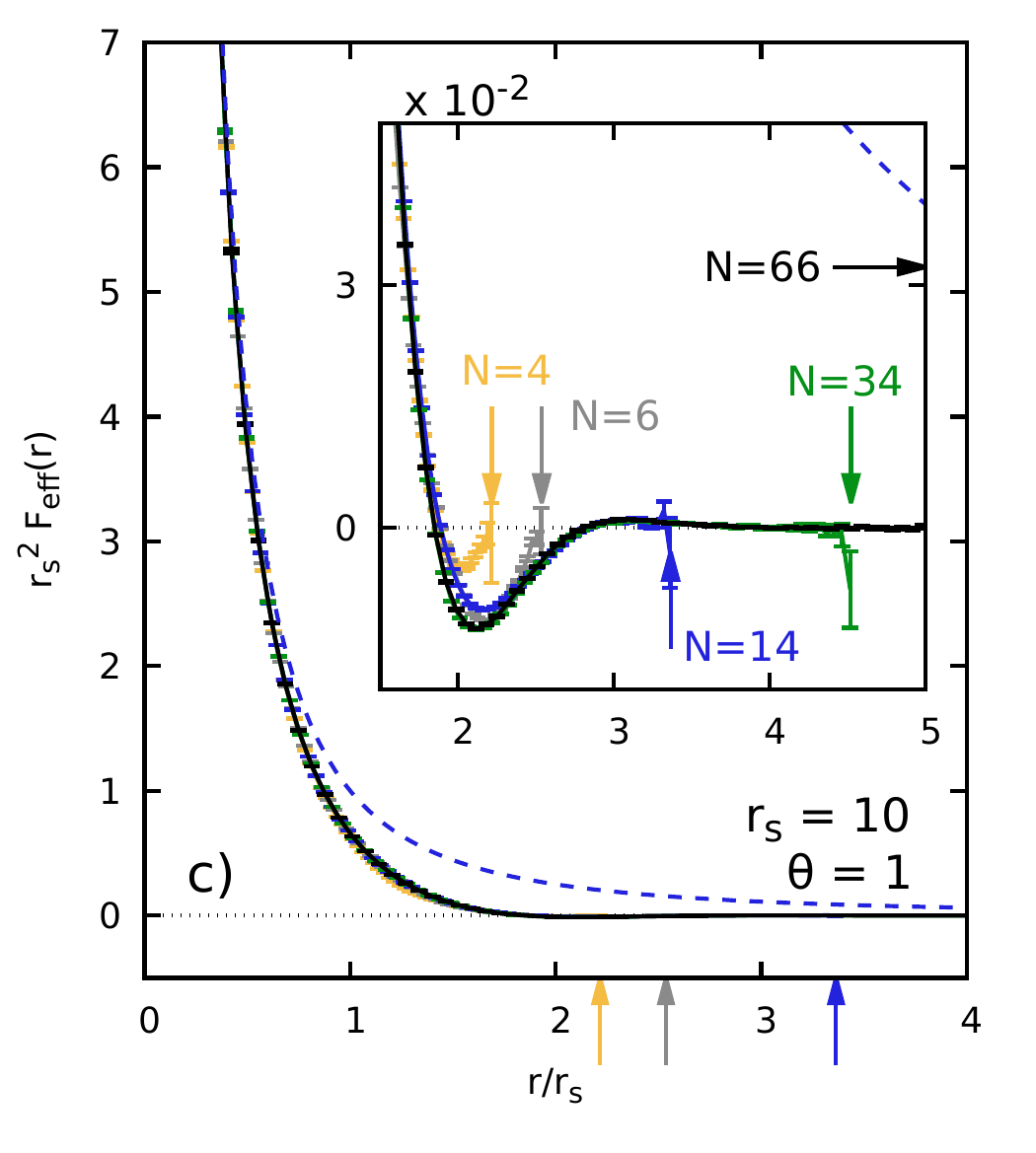}\includegraphics[width=0.475\textwidth]{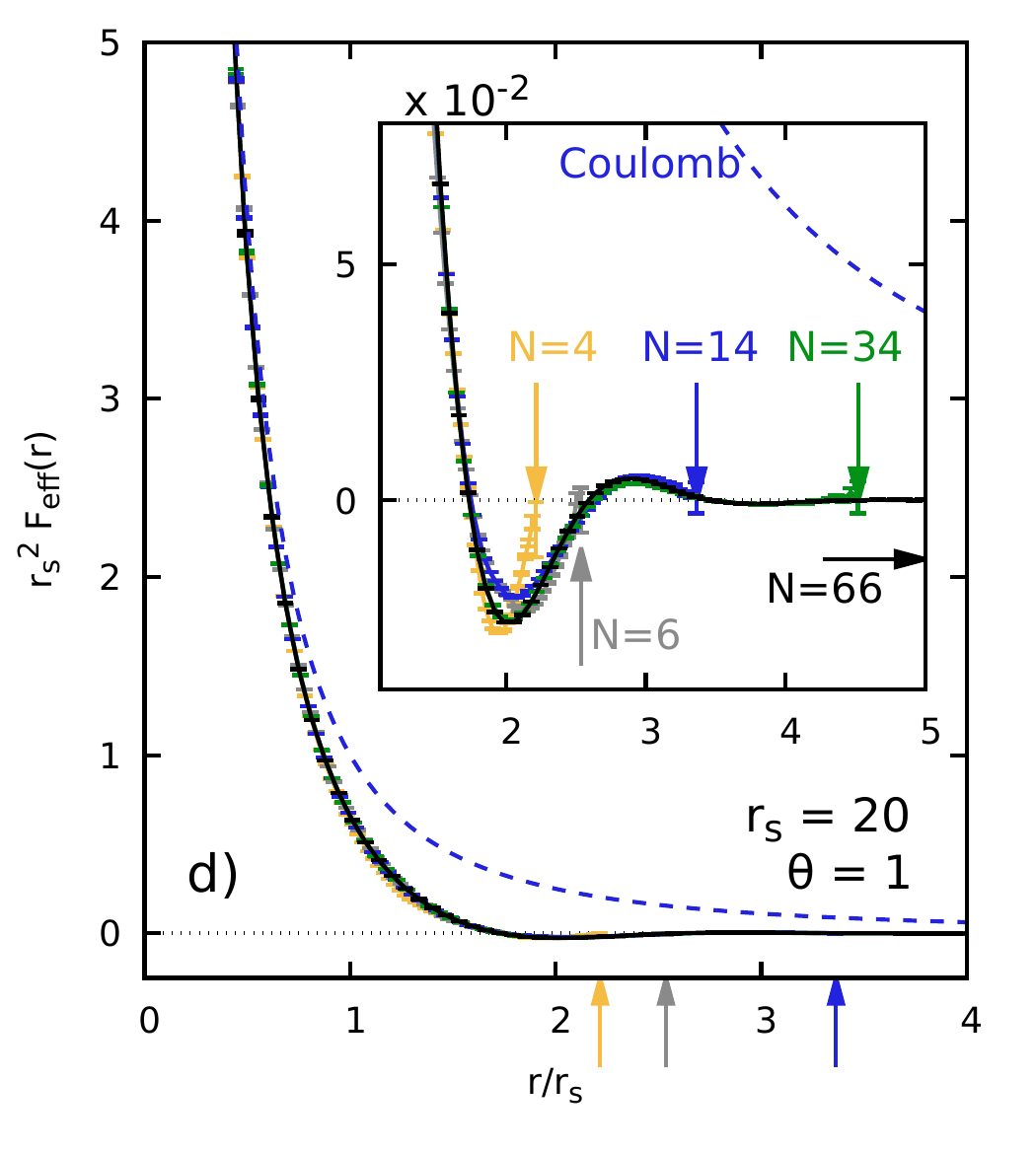}
\caption{\label{fig:FSC_theta1}
System-size dependence of PIMC results for the effective force $F_\textnormal{eff}(r)$ in the UEG at $\theta=1$ and a) $r_s=2$, b) $r_s=4$, c) $r_s=10$, and d) $r_s=20$. Symbols, yellow: $N=4$; grey: $N=6$; blue: $N=14$; red: $N=20$; green: $N=34$; black: $N=66$. Dashed blue curves: bare Coulomb force. The colored arrows indicate the maximum possible distance between to electrons in the simulation cell for different $N$. The insets show magnified segments.
}
\end{figure*} 

In the preceding sections, we have asserted that our PIMC simulations give us the exact solution to the full quantum many-body problem of interest. While factually true, this statement neglects an important point: PIMC simulations are carried out for a finite system size $N$. In contrast, practical applications require information about the thermodynamic limit, i.e., the limit of $N,\Omega\to\infty$ where the number density $n=N/\Omega$ is being kept constant. This discrepancy leads to the presence of so-called \emph{finite-size errors}, which must be taken into account~\cite{Fraser_PRB_1996,Drummons_PRB_2008,Holzmann_2011,Holzmann_PRB_2016,dornheim_prl,Chiesa_PRL_2006,dornheim_jcp,Dornheim_JCP_2021}.

In Fig.~\ref{fig:FSC_theta1}, we thus carefully analyze the dependence of our PIMC simulation data on the number of electrons $N$ for four different values of the density parameter $r_s$ at the electronic Fermi temperature, $\theta=1$. Let us begin our investigation at $r_s=2$, shown in Fig.~\ref{fig:FSC_theta1} a). This is a metallic density that can be probed experimentally for example in aluminum~\cite{Sperling_PRL_2015,aluminum1}. As a reference, we also include the bare Coulomb force $F_\textnormal{C}(r) = 1/r^2$ as the dashed blue curves in all panels. 

First and foremost, we notice that the effective force $F_\textnormal{eff}(r)$ decays remarkably fast with $r$ and does not exhibit the long Coulombic tail. This is a rather profound observation, which can hardly be overstated in its importance. In fact, it is well-known that the long-range nature of the Coulomb interaction can easily lead to a number of divergent terms in different theoretical approaches that require special care to be removed or avoided~\cite{quantum_theory,bonitz_book,krempbook}. Correspondingly, one might have assumed that \emph{ab initio} simulation methods would require one to perform simulations of mesoscopic system sizes such as $N\sim10^6$ to overcome these difficulties. In practice, this would have precluded the application of state-of-the-art methods like QMC or density functional theory in many cases. Fortunately, this \emph{long-range beast} is effectively tamed by the screening in charged many-body systems. It is important to note that this does not only apply to effective ion--ion interactions~\cite{Ramazanov_PRE_2015,zhandos1,quantum_theory} that are, on average, screened by the electronic medium, but does also occur in a purely electronic medium with a fixed, non-polarizable positive background. Therefore, the electrons are effectively \emph{near-sighted} and, in practice, do not experience correlations to electrons beyond the remarkably short range of $d\sim 3r_s$. We emphasize, though, that the above observation does not imply that the Coulomb interaction can be truncated at $d\sim 3r_s$ by introducing a sharp or shifted cut-off. In other words, the long range nature of the Coulomb pair interactions should be fully treated (see the Ewald summation for periodic systems) for the short range nature of the effective electronic interactions to properly emerge.

This empirical finding constitutes the basis for a number of applications in statistical physics, quantum chemistry, and related disciplines. In particular, one can use QMC methods (or less accurate, but computationally cheaper alternatives such as density functional theory) to accurately estimate exchange--correlation properties, that are of short-range nature, but cannot be estimated from any analytical theory. This information concerning the competition between Coulomb correlations, quantum degeneracy, and diffraction effects can then be combined with other methods to model the large scale behaviour of the system. The combination of computational methods with theoretical models like dielectric theories~\cite{stls_original,Tolias_JCP_2021,stls,stls2,tanaka_hnc} then allows for a highly accurate description of the system of interest everywhere~\cite{Chiesa_PRL_2006,Holzmann_2011,dornheim_prl,dornheim_cpp,Holzmann_PRB_2016,Dornheim_JCP_2021}. 
Likewise the fast decay of the effective force can be viewed as a theoretical justification for fragmentation methods~\cite{Y1991} that partition the electron density into smaller subsystems~\cite{EBCW2010}.

Let us now return to the task at hand, which concerns the assessment of finite-size errors in our PIMC data for the effective force $F_\textnormal{eff}(r)$. Evidently, any $N$-dependence in the different data sets is so remarkably small, that it can hardly be discerned with the naked eye. In addition, the vertical arrows at the $x$-axis show the maximum possible particle separation $r_\textnormal{max}^N$ for each $N$. Yet, $F_\textnormal{eff}(r)$ has nearly decayed to zero in all depicted cases, as seen particularly well in the inset that focuses on a magnified segment located around $r_s \leq r \leq 2.5r_s$. On this scale, we do observe small finite-size effects for both $N=4$ and $N=6$, with $F_\textnormal{eff}(r)$ decaying too fast especially for the former case.

\begin{figure*}\centering
\includegraphics[width=0.475\textwidth]{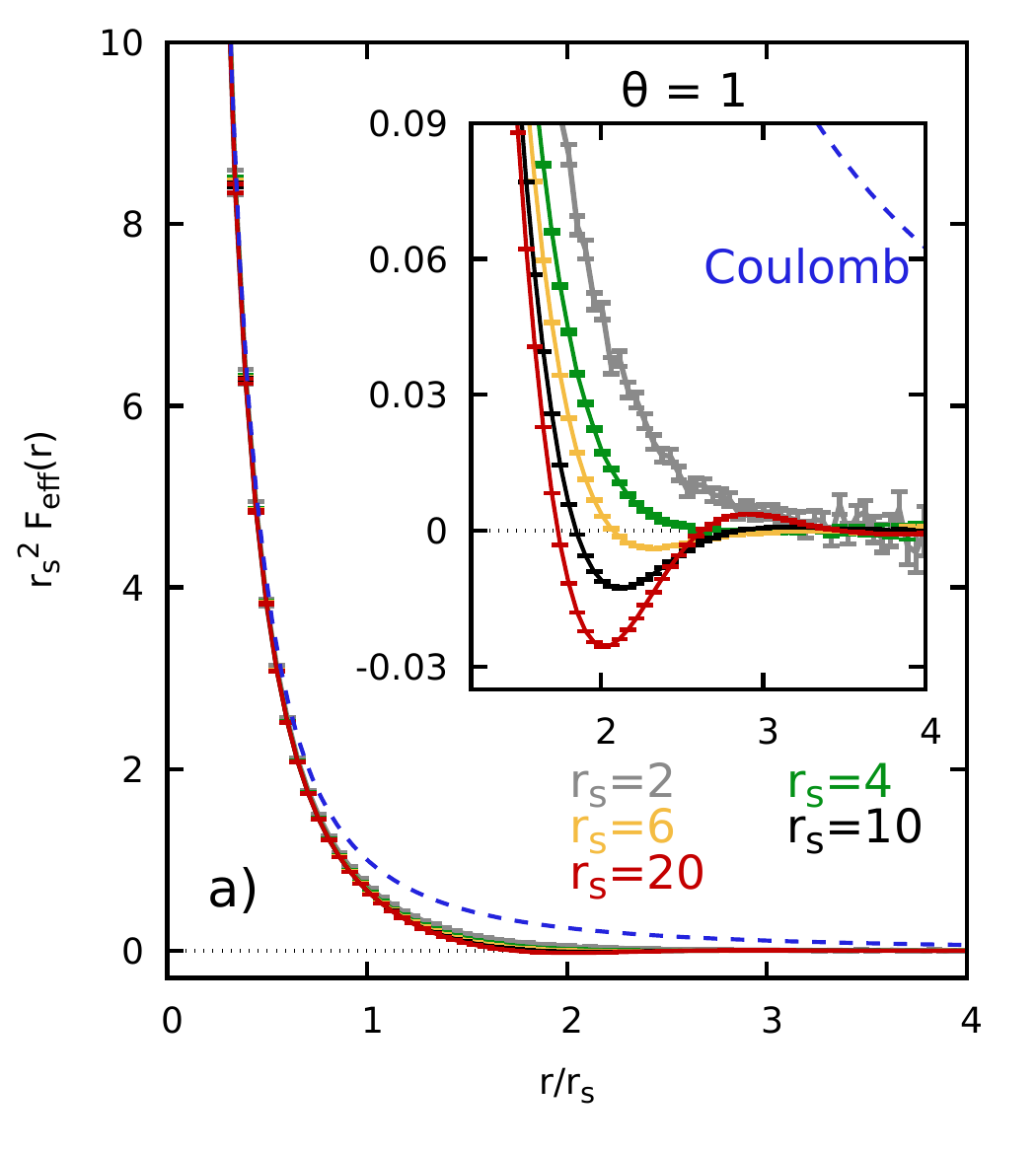}\includegraphics[width=0.475\textwidth]{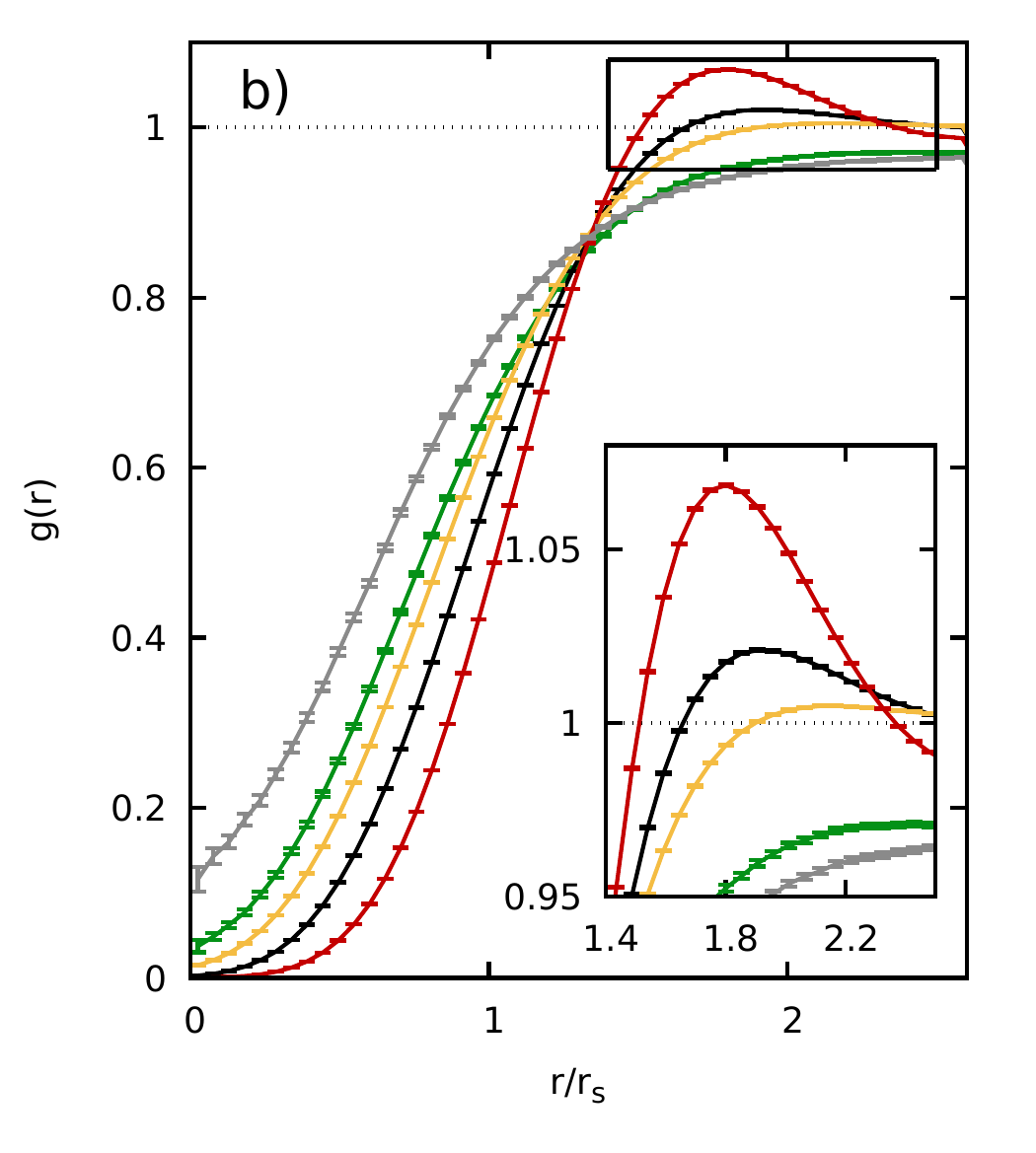}
\caption{\label{fig:Force_theta1_rs}
Left: dependence of the PIMC results for the effective force $F_\textnormal{eff}(r)$ at $\theta=1$ on the density parameter $r_s$. The inset shows a magnified segment around the attractive minimum for $r_s\gtrsim 6$. Right: corresponding PIMC results for the pair correlation function $g(r)$, with the inset showing a magnified segment around its peak.
}
\end{figure*} 

In panel b), we show analogous results for a lower density, $r_s=4$, which is close to sodium~\cite{Huotari_PRL_2010}. Overall, the behaviour of the PIMC data is very similar compared to the previous case of $r_s=2$. The main difference lies in the fact that finite-size effects for $N=4$ and $N=6$ are even smaller, with the $N=4$ data set exhibiting the correct asymptotic decay. This is a general trend, which is consistent to previous investigations of finite-size errors in the warm dense UEG~\cite{dornheim_prl,review,Dornheim_JCP_2021}, namely: as the density increases (\emph{i.e.}, decreasing the density parameter $r_s$), larger numbers of QMC simulated electrons are necessary to achieve convergence of wave-number resolved quantities such as the static structure factor $S(q)$. In the present work, we observe that this is a direct consequence of the fact that many-body correlations as they are embodied in $F_\textnormal{eff}(r)$ have a characteristic scale that exceeds the length of the simulation cell $L=\Omega^{1/3}$ in these cases.

We proceed with $r_s=10$ [panel c)], which corresponds to a moderately coupled system at the margin of the electron liquid regime~\cite{dornheim_electron_liquid}. Such conditions are very interesting for many reasons. \emph{Quantum liquids} allow one to gain new insights into the interplay of quantum effects with correlations and incipient localization effects. In addition, the UEG is known to exhibit a wealth of intriguing phenomena in this regime, such as the \emph{roton feature} in the dynamic structure factor $S(\mathbf{q},\omega)$~\cite{dornheim_dynamic,dynamic_folgepaper,Dornheim_Nature_2022}; see Sec.~\ref{sec:potential} for a detailed discussion of this point. First and foremost, we note that finite-size errors are almost non-existent over the entire distance range. In addition, we find an \emph{attractive minimum} in the effective force around $r\gtrsim 2 r_s$. While being comparably small in magnitude at these conditions, this effect is significant and clearly not an artifact due to the finite number of electrons in the simulation cell; indeed, only $N=4$ cannot correctly resolve this feature as its location exceeds the maximum possible inter-particle separation in this case. The physical origin of the attraction and its potential consequences for the physics of the UEG are discussed in Sec.~\ref{sec:dependence}.

Finally, we also show results for even stronger coupling, $r_s=20$ [panel d)]. In this case, the attractive minimum in $F_\textnormal{eff}(r)$ becomes even more pronounced, and is followed by a local maximum around $r=3r_s$. In fact, a very shallow yet significant second minimum can even be discerned around $r=4r_s$. These features are a direct consequence of the incipient mesoscopic order in the system upon entering the electron liquid regime~\cite{dornheim_electron_liquid,quantum_theory}, see also the discussion in Sec.~\ref{sec:classical}.

\subsection{Dependence on density and temperature\label{sec:dependence}}

\begin{figure*}\centering
\includegraphics[width=0.475\textwidth]{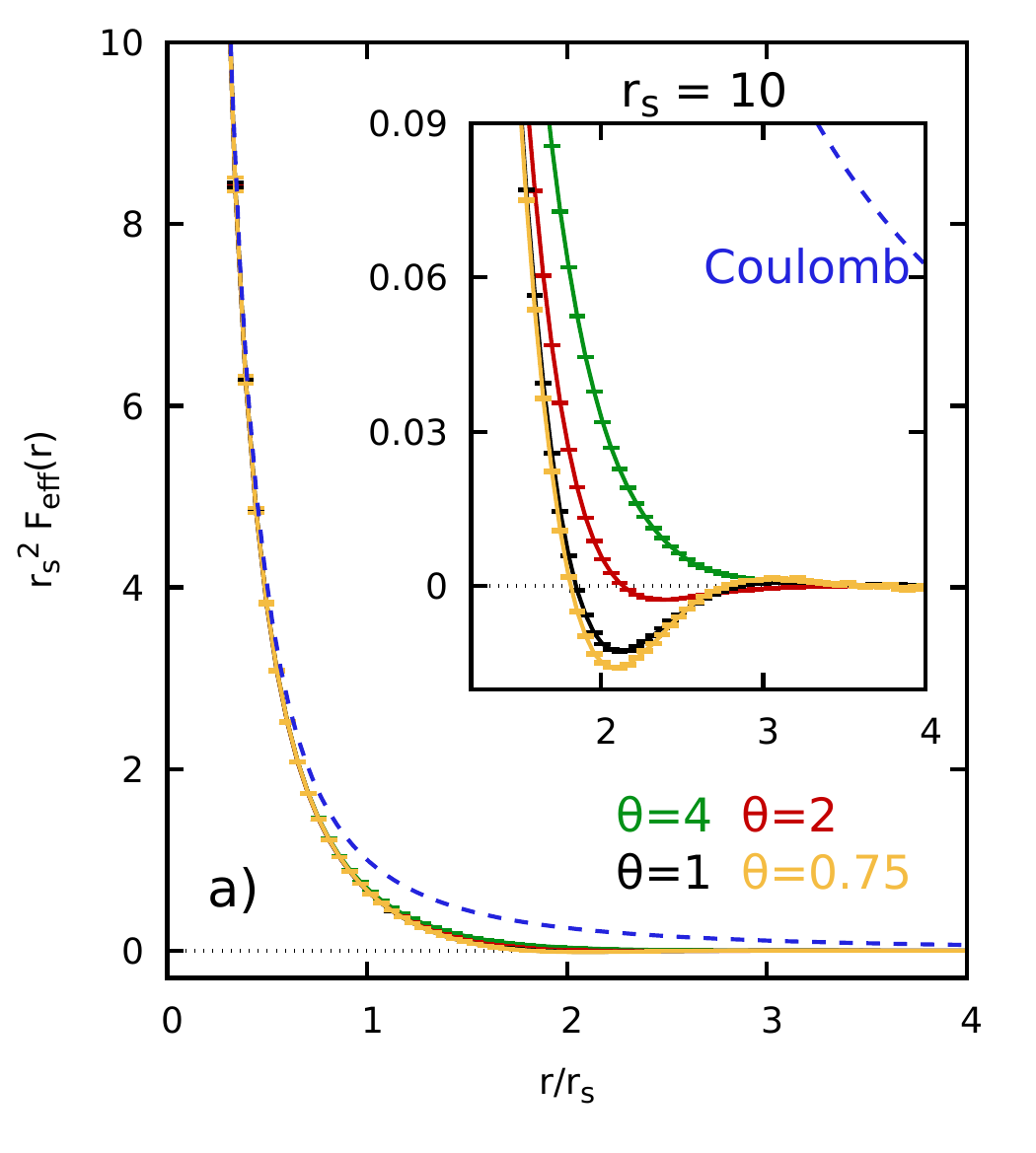}\includegraphics[width=0.475\textwidth]{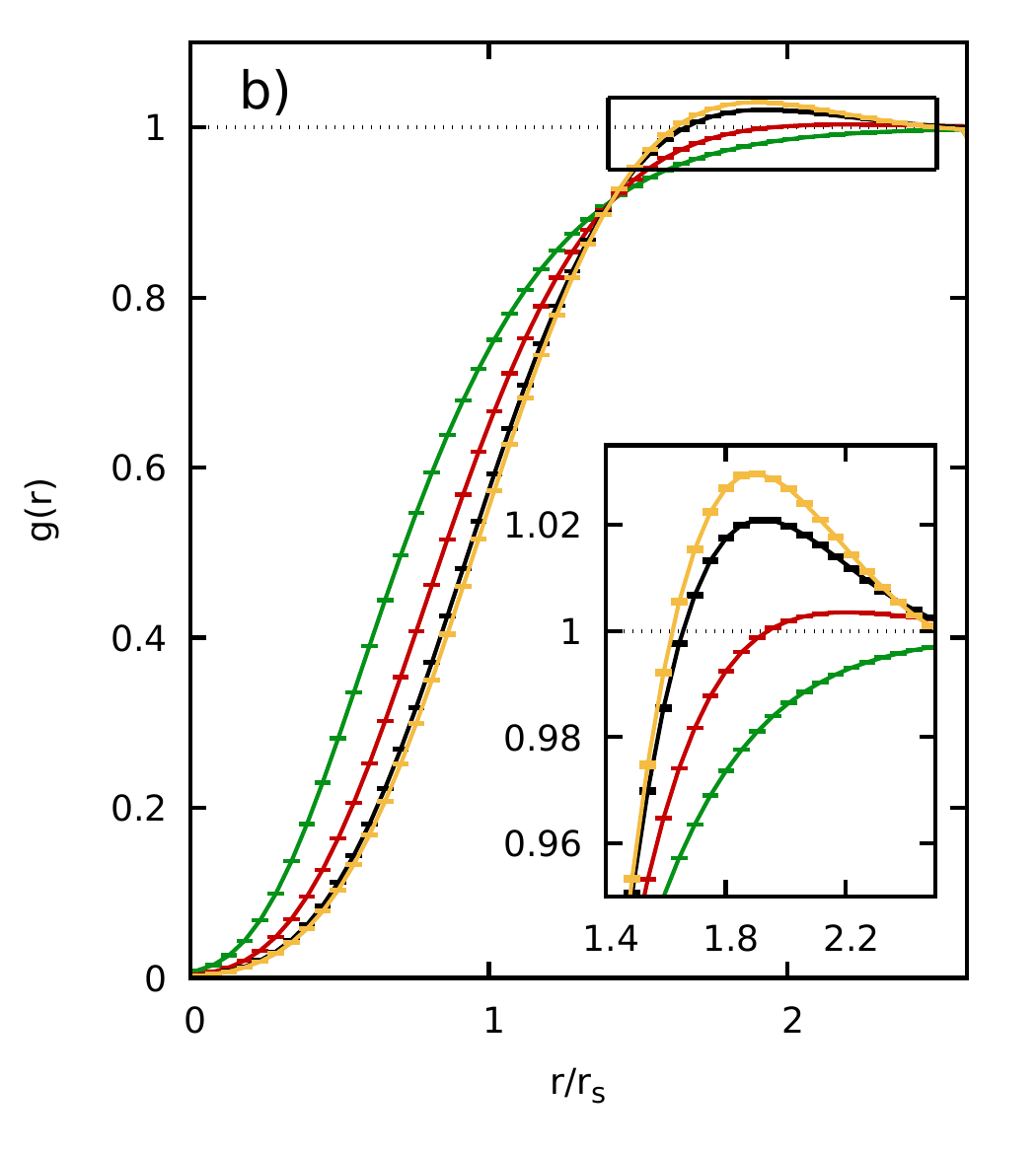}
\caption{\label{fig:Force_rs10_theta}
Left: dependence of the PIMC results for the effective force $F_\textnormal{eff}(r)$ at $r_s=10$ on the degeneracy temperature $\theta$. The inset shows a magnified segment around the attractive minimum for $\theta\lesssim2$. Right: corresponding PIMC results for the pair correlation function $g(r)$, with the inset showing a magnified segment around its peak.
}
\end{figure*} 

In what follows, we shall more closely examine the main physical trends in $F_\textnormal{eff}(r)$. To this end, we show PIMC simulation results for different values of $r_s$ at the electronic Fermi temperature $\theta=1$ in Fig.~\ref{fig:Force_theta1_rs} a). We note that we have re-scaled the forces by a factor of $r_s^2$ to make them comparable between different densities. Evidently, all depicted data sets exhibit a qualitatively very similar behaviour. At large distances, they quickly decay towards zero and do not exhibit the long-range Coulomb tail, as discussed at length in the previous section. Moreover, the effective force is dominated by Coulomb repulsion in the limit of small distances on the depicted scale. The main feature in Fig.~\ref{fig:Force_theta1_rs} is the density-dependence of the effective attraction between the electrons, which we find for $r_s\gtrsim 5$ in our PIMC results. This is a signature of short range order, as discussed in Sec.~\ref{sec:classical}. Naturally, electronic attraction suggests a kind of pairing mechanism such as the formation of \emph{Cooper pairs} in the BCS theory for superconductivity~\cite{Bardeen_PhysRev_1957}. Yet, in that case, the effective attraction is mediated by phonons, \emph{i.e.}, by the underlying ionic structure. In contrast, the rigid, uniform, and non-polarizable background of the UEG, by definition, cannot be the origin of this effect.
Instead, it is purely due to the electronic medium, and can be intuitively understood by considering the sketch in Fig.~\ref{fig:rectangle} shown above. Let us again examine the blue and green beads; clearly, the former is effectively \emph{pushed} towards the latter by the two red beads at the bottom. While in the depicted example the Coulomb repulsion between the green and blue beads still exceeds this \emph{push}, this need not be the case. In particular, the UEG undergoes an incipient localization when the density is decreased. As a direct consequence of the role of $r_s$ as the \emph{quantum coupling parameter}, the electrons get 1) more clearly separated and 2) more ordered. In other words, they are, on average, pushed by the medium towards a spatial structure that minimizes their free energy landscape.

\begin{figure*}\centering
\includegraphics[width=0.475\textwidth]{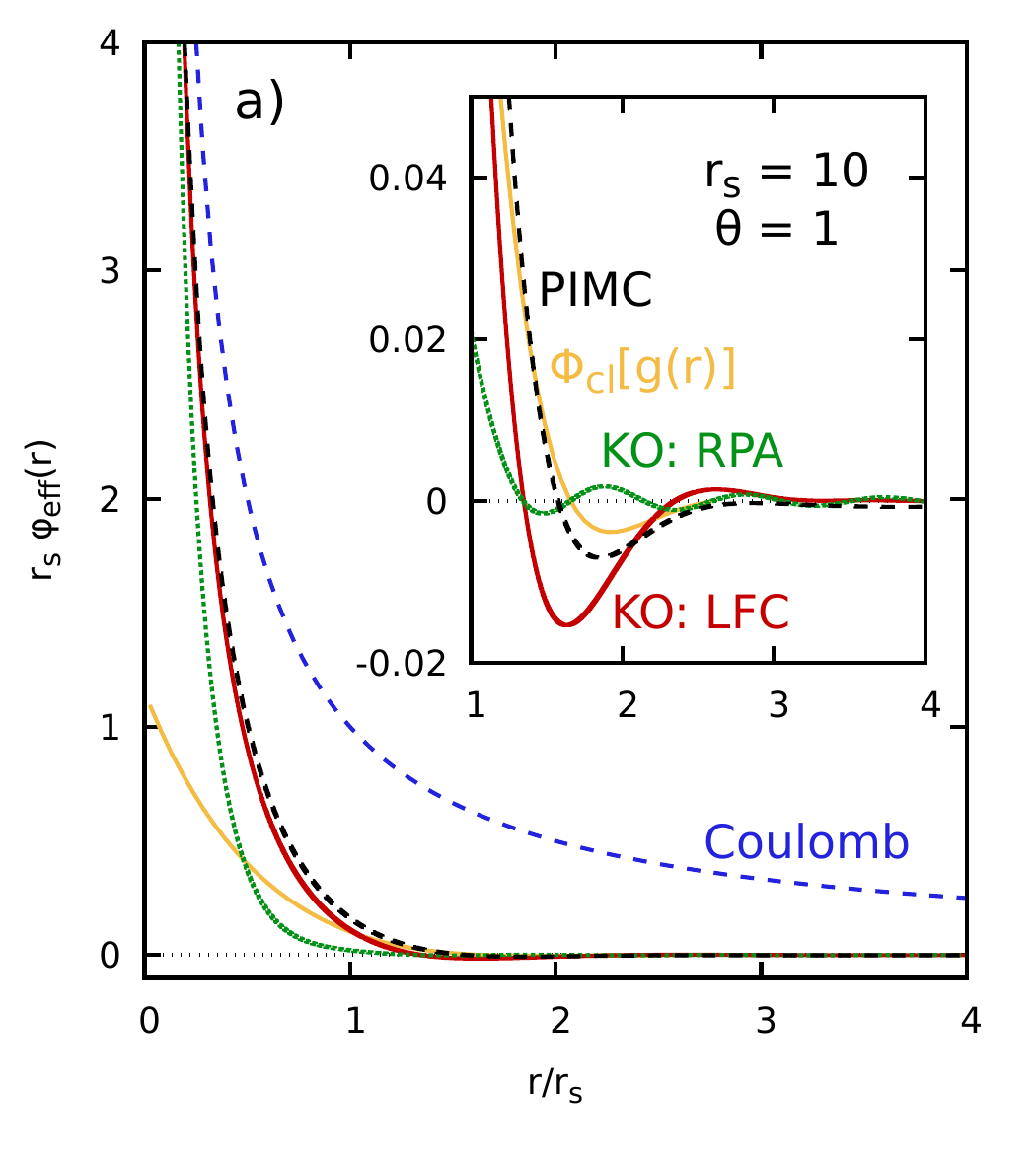}\includegraphics[width=0.475\textwidth]{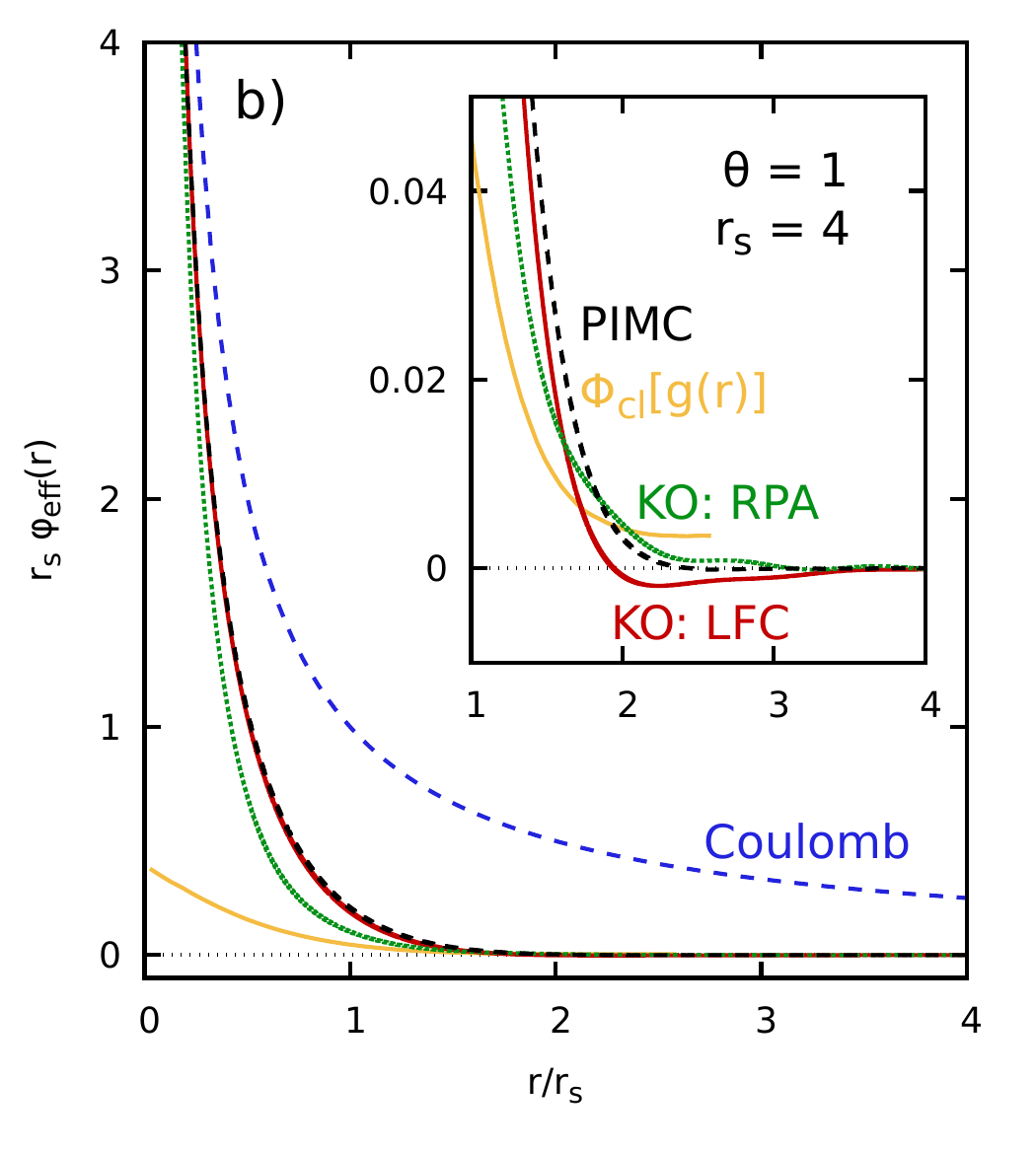}
\caption{\label{fig:potential}
Comparison of PIMC results for the effective potential $\phi_\textnormal{eff}(r)$ [Eq.~(\ref{eq:Phi})] (dashed black) to the KO potential using the SLFC from Ref.~\cite{dornheim_ML} (solid red), RPA (dotted green), the classical relation Eq.~(\ref{eq:phi_cl})] (solid yellow), and bare Coulomb (dashed blue) at $\theta=1$ for $r_s=10$ (left) and $r_s=4$ (right).}
\end{figure*}

To further illustrate this effect, we show PIMC results for the pair correlation function $g(r)$ in Fig.~\ref{fig:Force_theta1_rs} b). Evidently, $g(r)$ exhibits a (small) peak exactly for those $r_s$-values where we find the attractive feature in $F_\textnormal{eff}(r)$. For completeness, we note that the peak location is similar, but is not identical to the location of the latter, in contrast to the classical limit, where the $F_\textnormal{eff}(r)$ and $g(r)$ peak locations coincide, see Eq.~(\ref{eq:phi_cl}) and Fig.~\ref{fig:potential}.

Let us briefly touch upon the possible physical consequences of the effective electronic attraction in the UEG. Firstly, it constitutes a direct measure for the degree of UEG collective behaviour, which gives rise to various interesting phenomena such as the spectrum of density fluctuations~\cite{Kalman,Dornheim_Nature_2022}. In addition, it has been speculated that the attraction might constitute a possible pairing mechanism that would eventually lead to superconductivity in the UEG at $r_s\gtrsim 40$ and low temperatures~\cite{Takada_PRB_1993}. Similar possibilities have been suggested for semiconductors~\cite{Vignale_Singwi_PRB_1985} and metallic hydrogen~\cite{Richardson_1997a} in the literature. A detailed study of these conditions is beyond the scope of the present work and deserves dedicated exploration in the future.

A further interesting variable is the dependence of the attraction on the reduced temperature $\theta$, which we investigate in Fig.~\ref{fig:Force_rs10_theta}a) for $r_s=10$. Remarkably, the overall dependence of the effective electronic force in the UEG on the temperature is quite small. The most pronounced differences that we find concern the attractive feature around $r\sim 2r_s$. While the two curves for $\theta=0.75$ and $\theta=1$ are in very close agreement, the minimum becomes very shallow for $\theta=2$ and completely vanishes for $\theta=4$. This is again directly correlated to the presence of a peak in $g(r)$, which we show in Fig.~\ref{fig:Force_rs10_theta}b) at the same conditions.

\subsection{Comparison to theoretical models\label{sec:comparison}}

\begin{figure*}\centering
\includegraphics[width=0.475\textwidth]{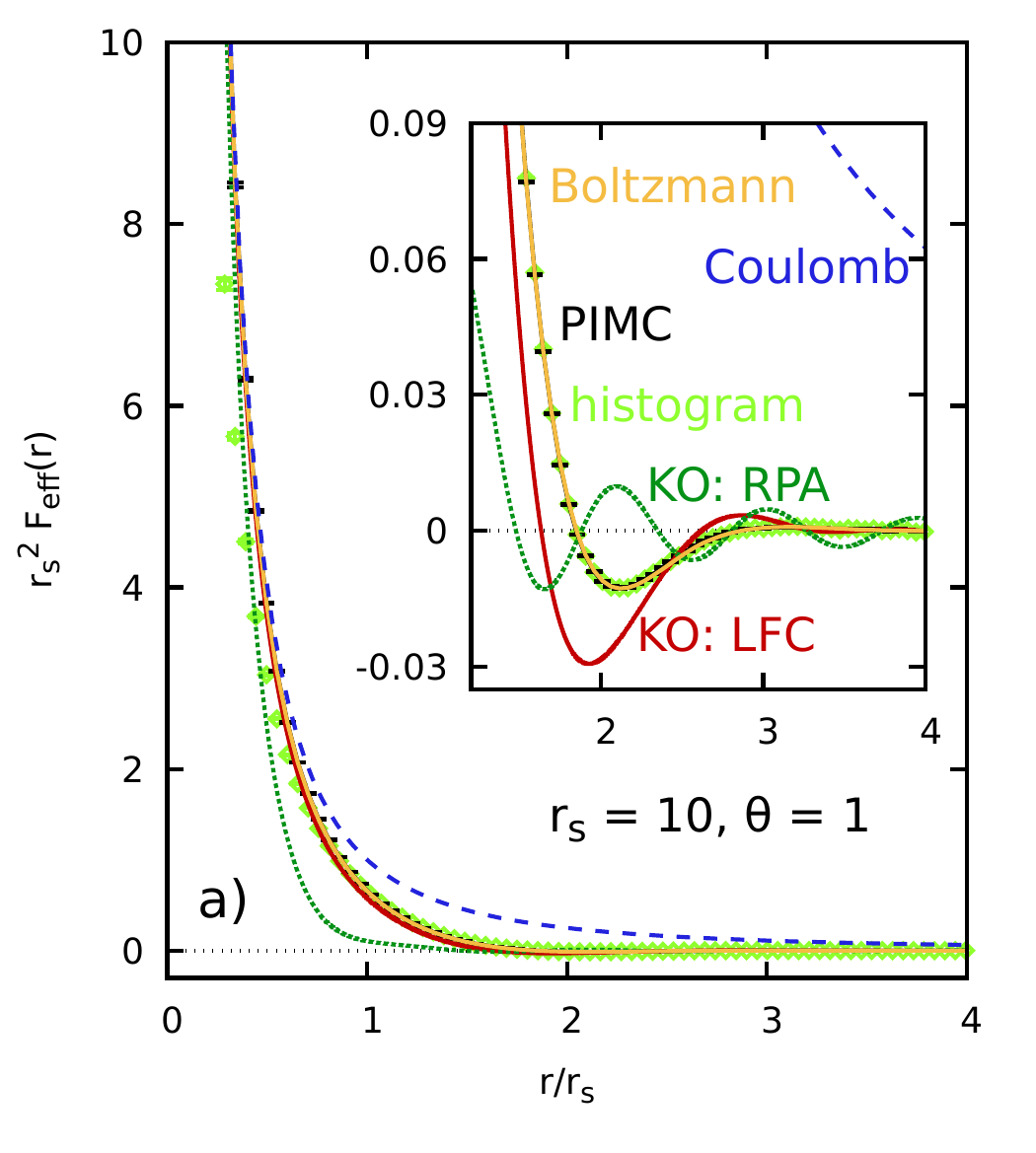}\includegraphics[width=0.475\textwidth]{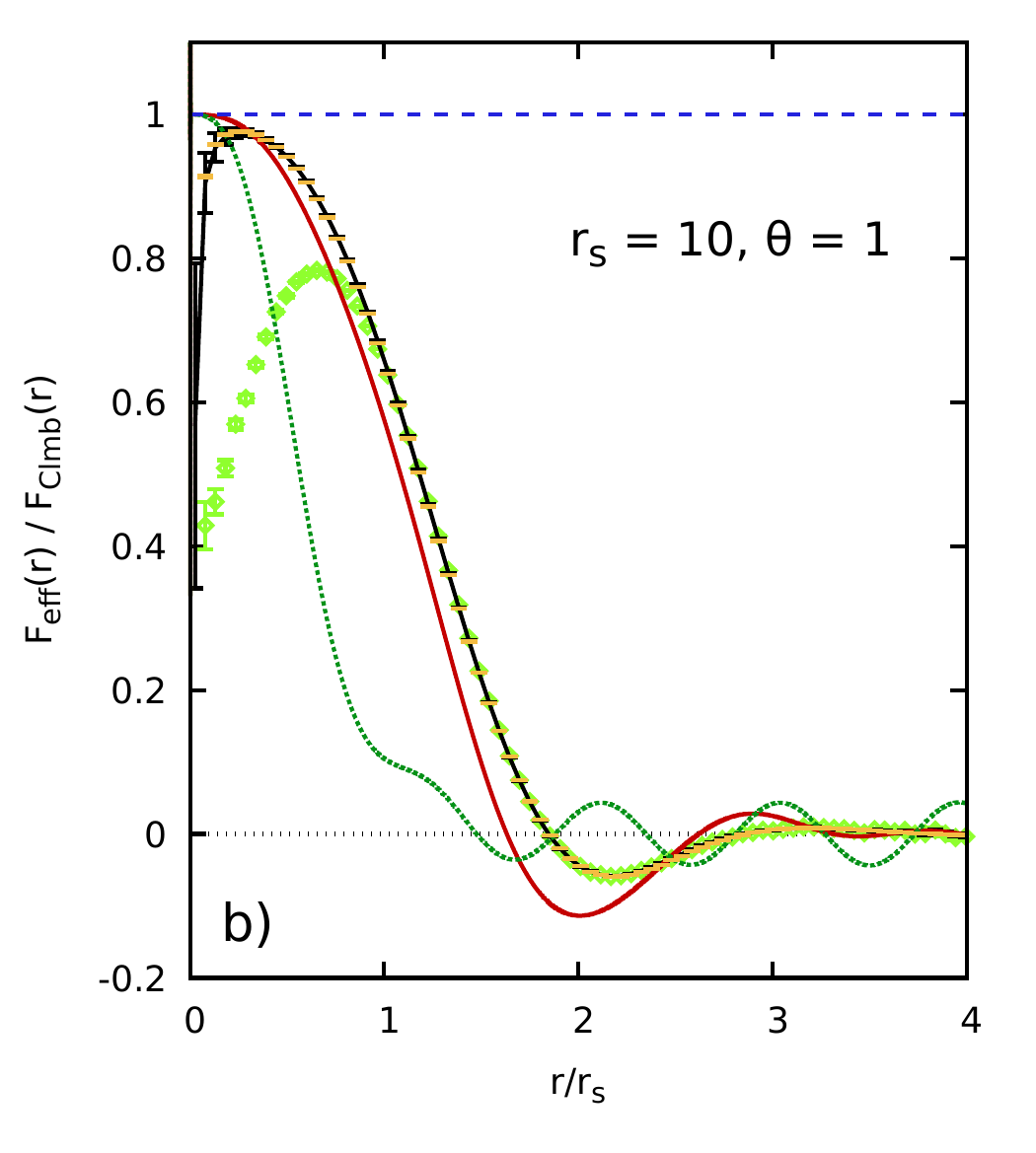}\\\includegraphics[width=0.475\textwidth]{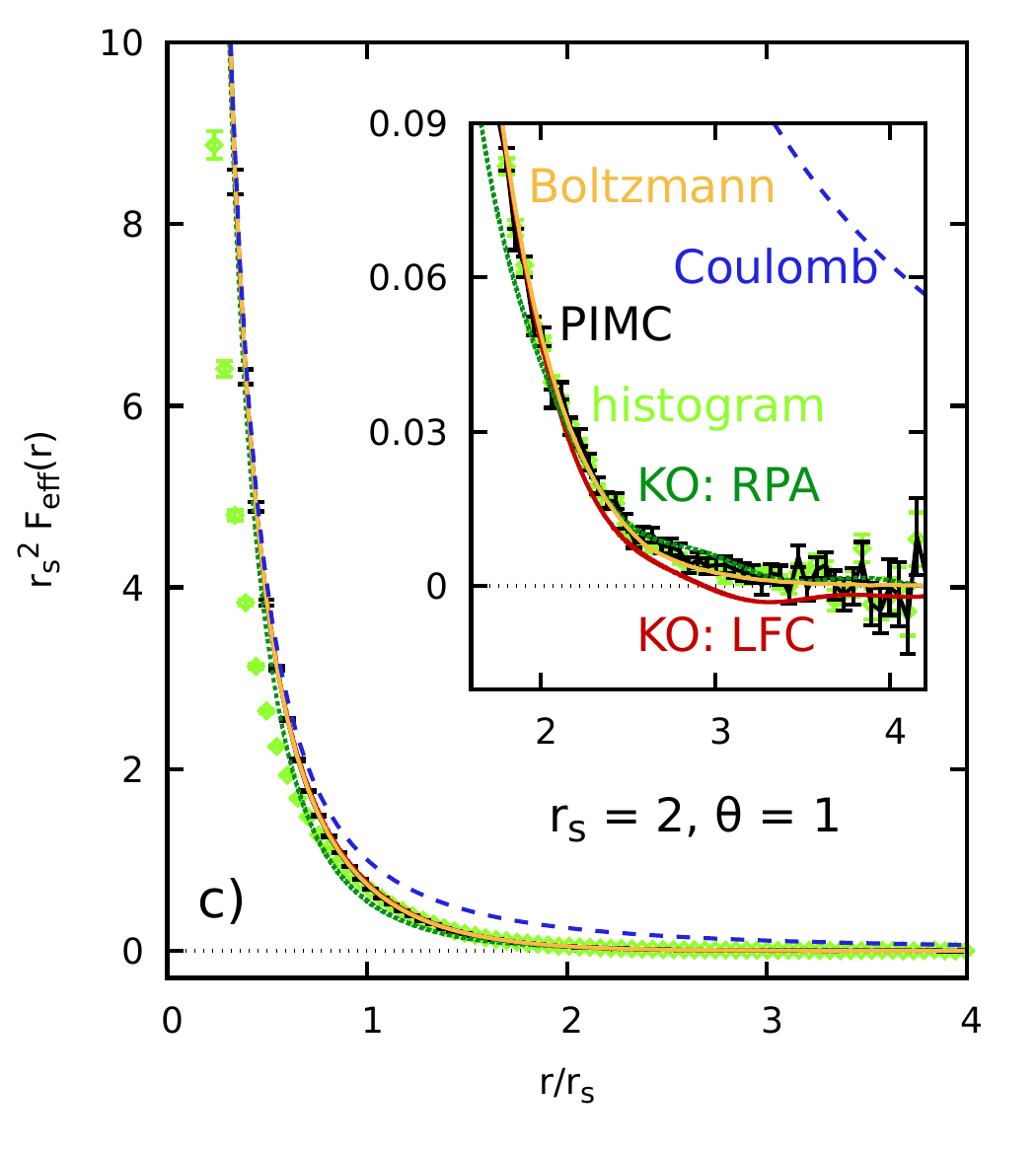}\includegraphics[width=0.475\textwidth]{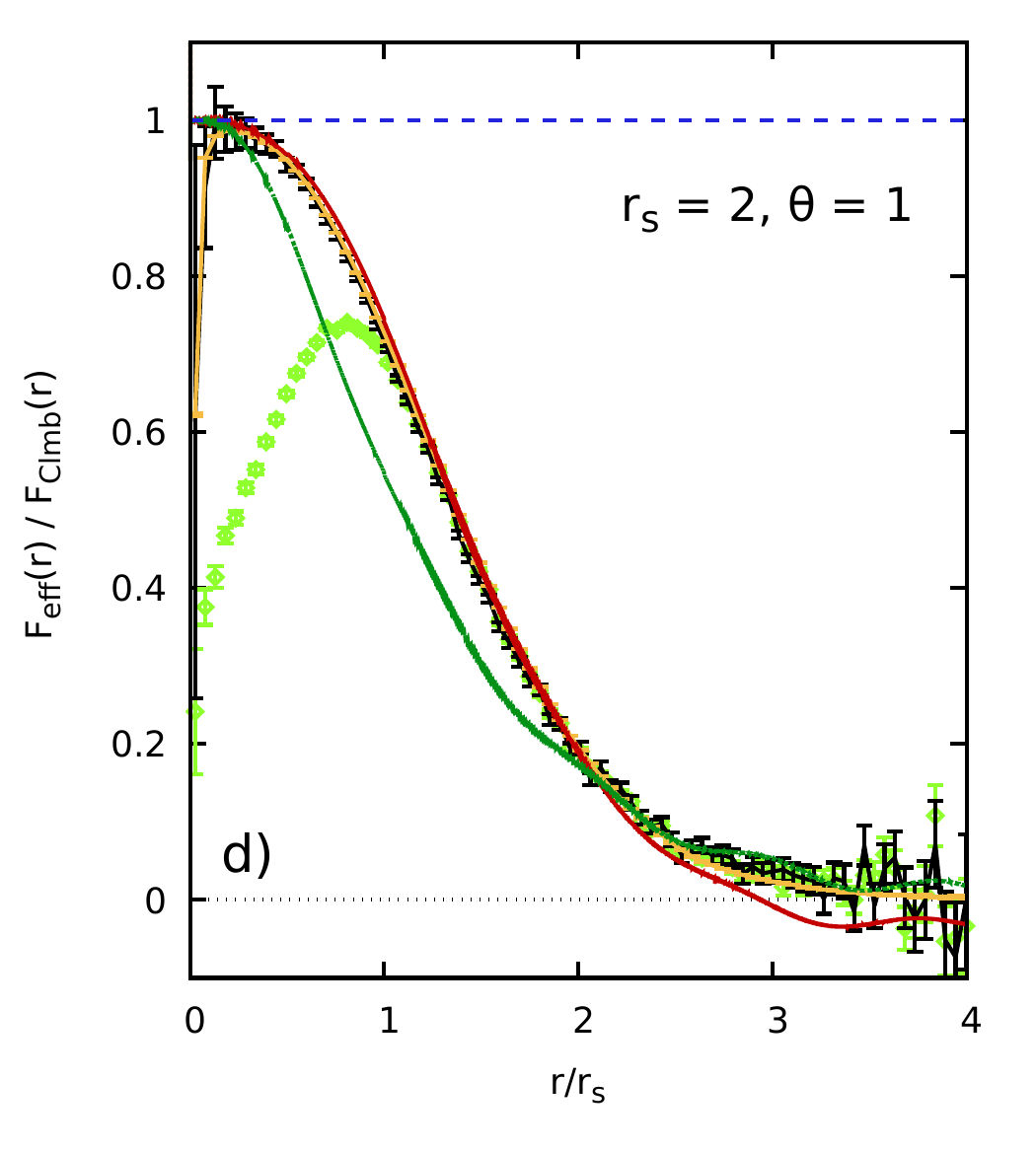}
\caption{\label{fig:compare}
Left column: Comparison of PIMC results for the effective force $F_\textnormal{eff}(r)$ to other theories at $\theta=1$ for $r_s=10$ (top) and $r_s=2$ (bottom). Black: exact PIMC, Eq.~(\ref{eq:rigorous}); yellow: Boltzmann PIMC (see text); light green: naive histogram estimator; solid red (dotted green): Kukkonen Overhauser (KO), Eq.~(\ref{eq:KO}), using as input the static LFC from Ref.~\cite{dornheim_ML} (within RPA, i.e., $G(q)\equiv 0$); dashed blue: bare Coulomb force: Right column: Effective forces relative to the bare Coulomb force.
}
\end{figure*} 

A further interesting topic of investigation is the capability of theoretical models to capture the microscopic effective interaction between two electrons in the UEG. This is explored in Fig.~\ref{fig:compare}. Specifically, the left column shows $F_\textnormal{eff}(r)$ from different theories, and the top row corresponds to $r_s=10$ and $\theta=1$. As usual, the dashed blue line shows the bare Coulomb force that has been included as a reference, and the black symbols depict our \emph{quasi-exact} PIMC data [cf.~Eq.~(\ref{eq:rigorous})] with their corresponding error bars. In addition, the dotted green line shows the linear-response prediction within the RPA, which has been obtained from Eqs.~(\ref{eq:KO}-\ref{eq:F_KO}) by setting $G(q)\equiv0$. Evidently, the RPA converges towards Coulomb at small distances. This can be seen particularly well in Fig.~\ref{fig:compare} b), where the ratio of the respective data sets to the bare Coulomb force has been plotted. Interestingly, we observe that the exact PIMC results start to systematically deviate from the Coulomb force in the limit of very small distances. For completeness, it is worth pointing out that the increasing error bars in the PIMC data towards contact are a direct consequence of the histogram estimation of $F_\textnormal{eff}(r)$, since pair-particle encounters become exceedingly infrequent within our simulations as the distance decreases. In principle, this bottleneck can be overcome in future works via the implementation of umbrella sampling techniques that were originally introduced in classical Monte Carlo methods to improve the sampling of ultra-rare configurations~\cite{Torrie1077,Kastner2011}. The RPA substantially overestimates the drop in $F_\textnormal{eff}(r)$ due to the medium for $r\sim r_s$, and it exhibits Friedel oscillations~\cite{Dharma-wardana_POP_2022} for $r\gtrsim 2r_s$ that slowly decay with increasing $r$. We note that these oscillations are not a real feature of the UEG at the present conditions, but are a direct consequence of the insufficient treatment of electronic exchange--correlation effects within the RPA. 

In contrast, the solid red curve has been obtained by evaluating the KO potential~\cite{Kukkonen_PRB_1979,quantum_theory} with input from the neural-net representation of the static LFC $G(q;r_s,\theta)$ from Ref.~\cite{dornheim_ML}, which is based on accurate PIMC data. We note that the recently introduced analytical representation of the SLFC~\cite{Dornheim_PRB_ESA_2021} within the effective static approximation~\cite{Dornheim_PRL_2020_ESA} would basically lead to indistinguishable results. In practice, we find that the red curve constitutes a substantial improvement over the RPA, and exhibits the correct qualitative decay towards zero for $r\gtrsim r_s$. At the same time, qualitative differences arise which deserve a further exploration. In particular, the KO expression overestimates the magnitude of the effective electronic attraction at $r\sim2 r_s$, which can be roughly understood as follows. The minimum in $F_\textnormal{eff}(r)$ is a direct consequence of many-body effects, and the increasing impact of Coulomb correlations. Specifically, two particles are effectively pushed together by the surrounding medium. Yet, it is well known that the usual linear response does only explicitly include two-body correlations (although all orders are implicitly present in the SLFC), since three-body correlations are connected to the quadratic density response function, and so on~\cite{Dornheim_JPSJ_2021}. Therefore, we conclude that the linear-response description within the framework of KO cannot fully capture the effective attraction, which would require the incorporation of nonlinear effects~\cite{Dornheim_PRL_2020,Dornheim_PRR_2021,Dornheim_JCP_ITCF_2021}, see also the appendix for the analysis of the classical limit. In addition, the KO force monotonically converges towards the bare Coulomb force in the limit of $r\to0$ [see also Fig.~\ref{fig:compare} b)], and does not exhibit the relative drop of the exact PIMC results. 

To acquire additional insights into the physical mechanism behind the effective force, we carried out PIMC simulations using Boltzmann statistics. More specifically, these calculations include the full information about quantum diffraction and related effects, but the particles are assumed to be distinguishable; the anti-symmetry of the true fermionic density matrix under the exchange of particle coordinates is completely omitted. The results are shown as the yellow curve in Fig.~\ref{fig:compare}, and are in excellent agreement with the true PIMC data over the entire $r$-range.
We stress that this is a highly counter-intuitive finding, which calls for an explanation. In particular, it is well understood that fermionic exchange effects should substantially influence the behaviour of two electrons at sufficiently small distances $r$ at the present conditions. Indeed, if that was not be the case, there would be no need for fermionic PIMC simulations, and the notorious fermion sign problem would have been effectively circumnavigated. 

To understand this remarkable behaviour of $F_\textnormal{eff}(r)$, we need to recall its definition, and the corresponding estimation within the PIMC method discussed in Sec.~\ref{sec:medium} above. Specifically, Eqs.~(\ref{eq:rigorous2}) and (\ref{eq:rigorous}) imply that the effective force \emph{does not include} exchange and correlation effects between the two electrons themselves; this has been ensured by dividing by the expectation value of the corresponding normalization. In other words, our estimator corrects for fermionic exchange effects, which leads to the observed equality between fermions and Boltzmann particles. Aiming to further illustrate this point, we also implemented the naive histogram estimator discussed in the beginning of Sec.~\ref{sec:medium}. In this way, we only correct for correlation effects in the fermionic simulation, but do not take into account the impact of the fermionic antisymmetry under the exchange of particle coordinates.

The results are included as the light green diamonds in Fig.~\ref{fig:compare}. Evidently, they are in excellent agreement to the exact PIMC and Boltzmann PIMC results for $r\gtrsim r_s$, but strongly deviate for smaller $r$. In other words, fermionic exchange effects between two electrons only have an impact at very small distances, but do not significantly influence other quantum effects such as diffraction. 
In practice, the Pauli blocking prevents two identical fermions from occupying the same position, and can be interpreted as an additional effective force that separates two particles (i.e., the "degeneracy pressure"). Yet, this effective repulsion does not directly show up in the histrogram estimator for $F_\textnormal{eff}(r)$. Indeed, we use the PIMC method to compute the expectation value of Eq.~(\ref{eq:F}), where we only sum over the Ewald forces. In contrast, the Pauli repulsion manifests indirectly due to the cancellation of positive and negative terms during the PIMC simulation~\cite{dornheim_sign_problem,Dornheim_JPA_2021}. In practice, this leads to the following: holding together two identical fermions to short distances $r<r_s$, which is the only case when we sample $F_\textnormal{eff}(r<r_s)$ in our PIMC simulations, requires the medium to counteract the effective Pauli repulsion. Yet, the latter does not directly manifest in $F_\textnormal{eff}(r)$; the counter-acting \emph{push} bringing the electrons together, on the other hand, leads to the observed reduction in the light green curve. On the other hand, the consistent definition given by Eq.~(\ref{eq:rigorous}) completely removes this effect.

Let us next discuss the bottom row of Fig.~\ref{fig:compare}, where we show the same information for the metallic density of $r_s=4$. In this case, RPA is considerably more accurate compared to $r_s=10$ as electronic exchange--correlation effects are less important. In addition, we find smaller differences between the full PIMC data (black curve) and the KO results over the entire $r$-range. This, too, follows from the reduced importance of correlations, as nonlinear effects are directly related to higher-order correlation functions known from many-body theory~\cite{Dornheim_JPSJ_2021}. Finally, we note that the discrepancy between the exact PIMC results and the histogram estimator for small distances is actually increased compared to $r_s=10$, as fermionic exchange effects are more pronounced at the higher density~\cite{review}.

\begin{figure}\centering
\includegraphics[width=0.475\textwidth]{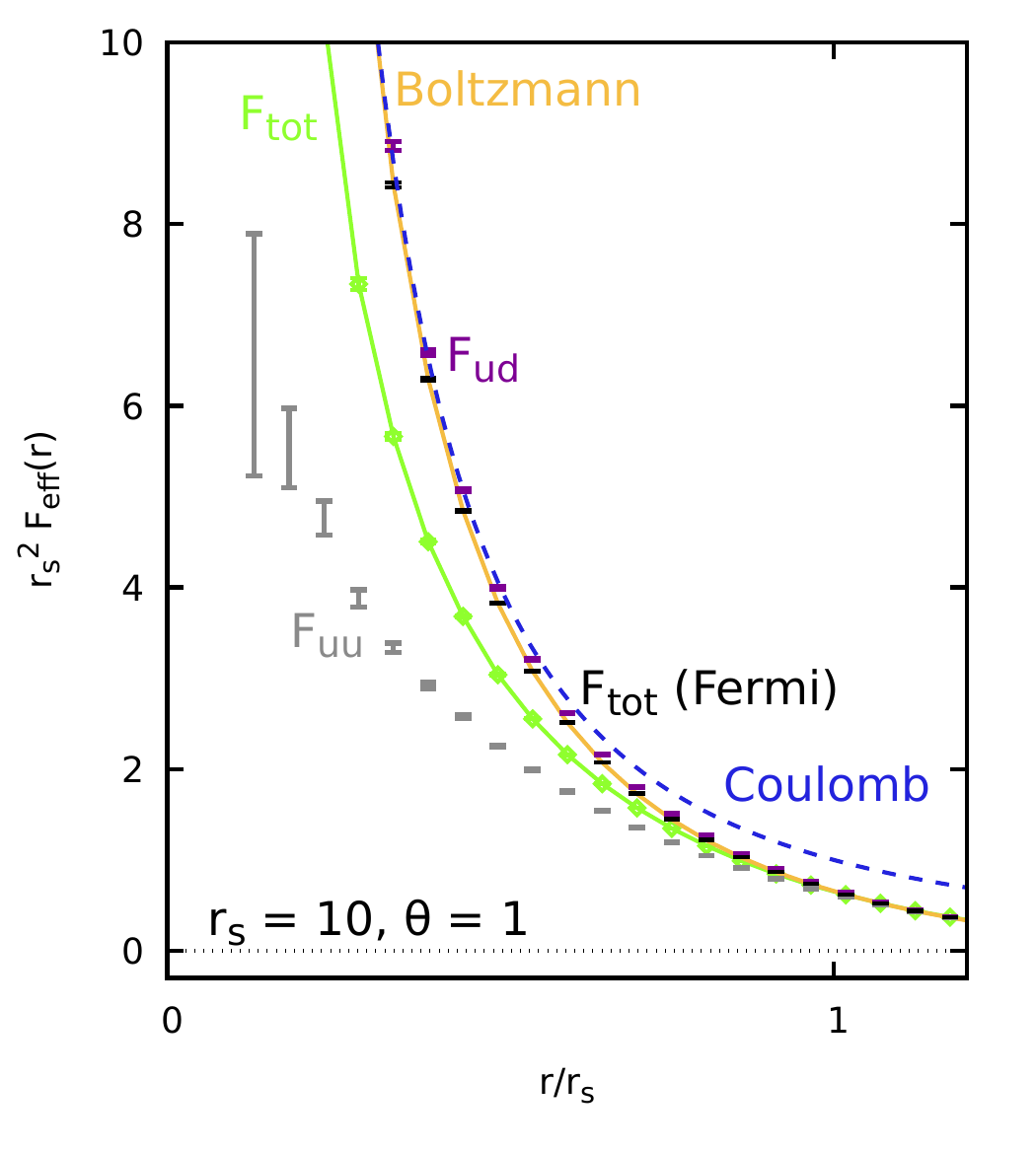}
\caption{\label{fig:spin}
Spin-resolved PIMC results for the effective force $F_\textnormal{eff}(r)$ at $\theta=1$, $r_s=10$, \emph{i.e.} for the same conditions as in Fig.~\ref{fig:compare} (top). Black: fermionic force $F_\textnormal{tot}(r)$ [cf.~Eq.~(\ref{eq:rigorous})]; light green: histrogram estimator for $F_\textnormal{tot}(r)$ not correcting for exchange; grey and purple: corresponding histogram estimators for $F_\textnormal{uu}(r)$ and $F_\textnormal{ud}(r)$; yellow: Boltzmann PIMC; dashed blue: bare Coulomb force.
}
\end{figure} 

We shall continue this investigation by exploring the \emph{spin-resolved} effective force shown in Fig.~\ref{fig:spin}. As usual, the black, yellow, and dashed blue curves show the full fermionic PIMC results [Eq.~(\ref{eq:rigorous})], the Boltzmann PIMC results, and the bare Coulomb force, respectively. In addition, the light green, grey and purple symbols show the effective force computed using the naive histogram estimator between any two electrons, between two electrons of the same spin-orientation (e.g. up-up) and between two electrons of opposite spin (up-down). Evidently, the latter closely follows the exact PIMC and Boltzmann PIMC results, as it is not directly influenced by exchange effects. The up-up force, on the other hand, is strongly reduced compared to the other data sets due to the counter-acting of the effective Pauli repulsion discussed earlier. For completeness, we note that no significant spin-dependence can be resolved in our spin-resolved implementation of the true fermionic expectation value Eq.~(\ref{eq:rigorous}), which is not pictured in Fig.~\ref{fig:spin}.

From a practical perspective, the observed trends open up intriguing new possibilities for the description of Fermi systems, in general, and many-electron systems, in particular. As we have mentioned above, the main bottleneck of QMC simulations is the exponential increase in the compute time with the system-size $N$. The short-range nature of the effective force between two electrons leads to a great reduction of this problem, as many-body correlation effects are basically limited to the range of $r\lesssim 3r_s$ in the regime of moderate coupling, $r_s\sim 1 - 10$. While this makes QMC studies feasible in many cases, a considerable part of the interesting density-temperature range remains out of reach~\cite{Yilmaz_JCP_2020,dornheim_POP}. Our present findings clearly indicate, that the impact of fermionic exchange-effects, which is the most challenging part for all QMC methods, is limited to even shorter length scales, and becomes negligible for $r\gtrsim r_s$. In a certain sense, this finding constitutes an empirical confirmation of Kohn's \emph{principle of nearsightedness}~\cite{Kohn_PNAS_2005} that has been introduced in the context of density functional theory~\cite{Kohn_PRL_1996}.
In particular, it opens up the enticing possibility for a further decomposition: specifically, we propose to use exact fermionic QMC simulations of very small systems to accurately capture the full interplay of Coulomb correlations with fermionic exchange effects for $r\lesssim r_s$; the still formidable effects of quantum diffraction are fully captured by the Boltzmann PIMC simulations, which do not suffer from the exponential bottleneck. Indeed, simulations with $N\sim10^2-10^3$ \emph{distinguishable electrons} are feasible; these information can finally be combined with dielectric theories such as the RPA to get an accurate description of the system over all length scales. The practical implementation of this idea will be explored in detail in a future publication.

\begin{figure*}\centering
\includegraphics[width=0.475\textwidth]{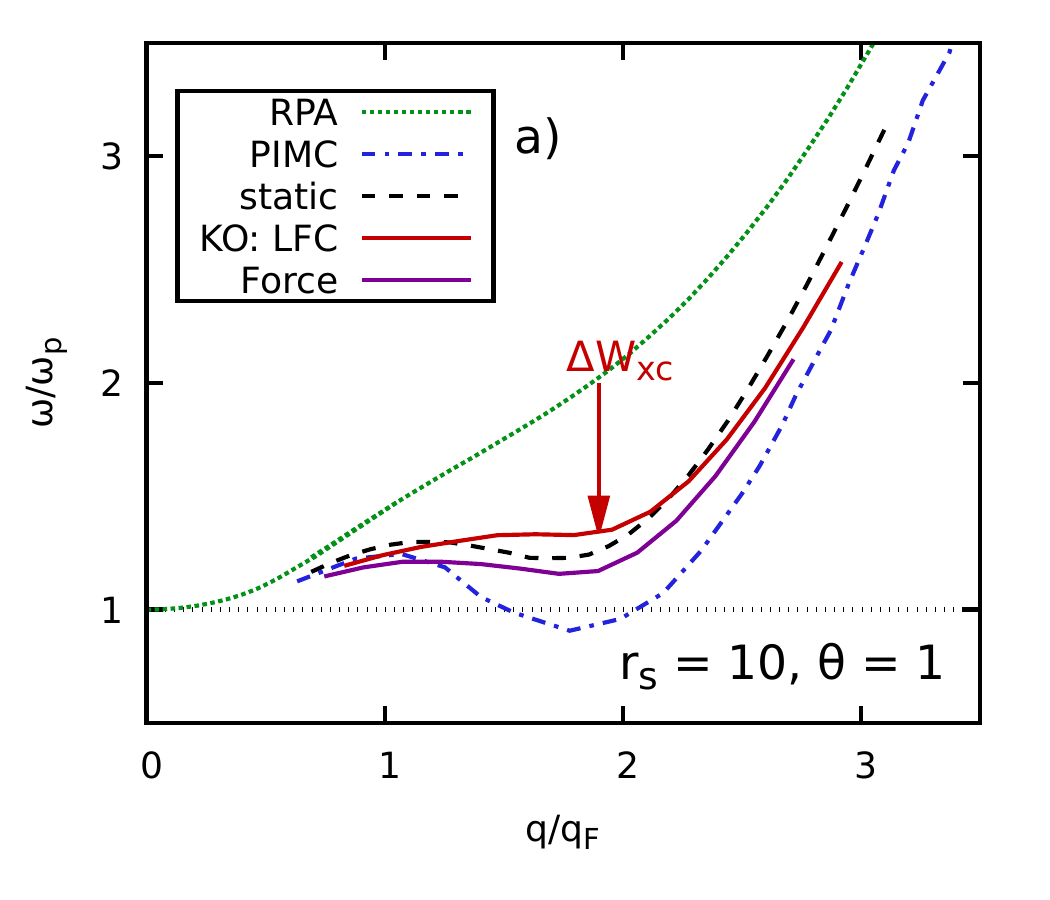}\includegraphics[width=0.475\textwidth]{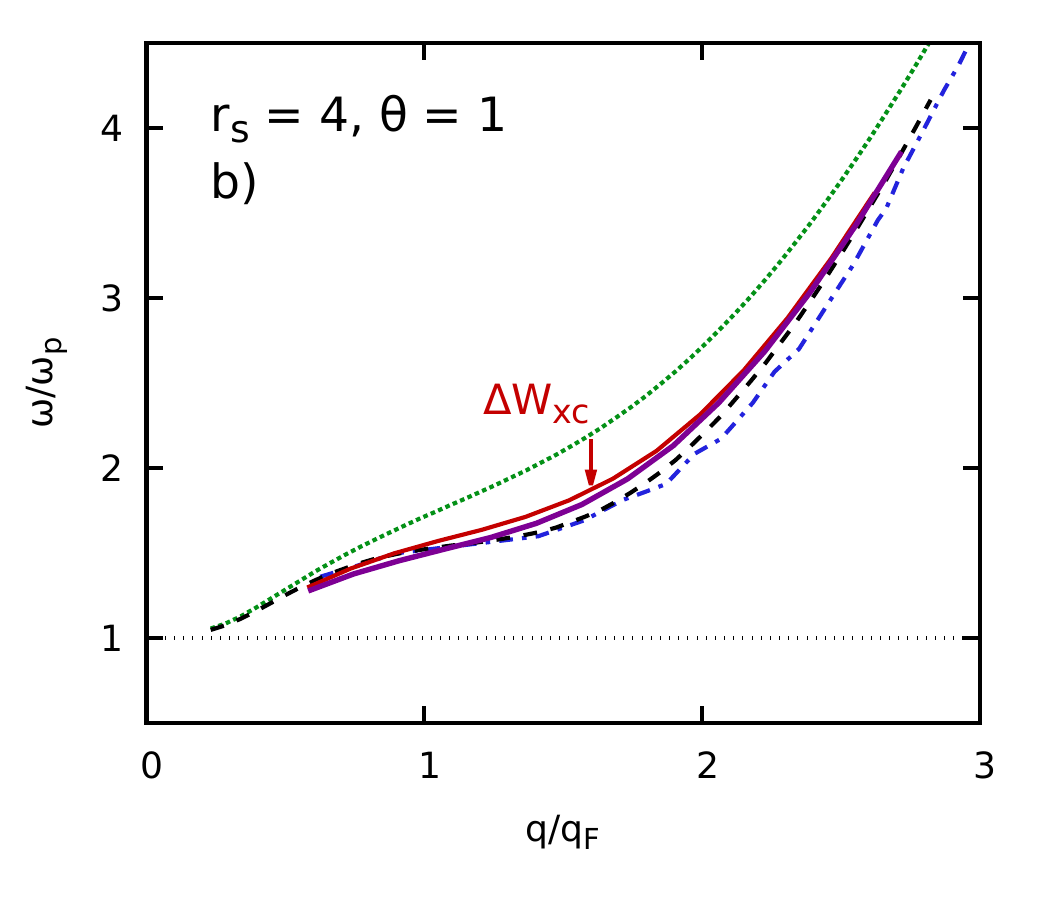}
\caption{\label{fig:dispersion} Spectrum of density fluctuation [estimated from the peak in the dynamic structure factor $S(q,\omega)$] in the UEG at $\theta=1$ for a) $r_s=10$ and b) $r_s=4$. Dotted green: RPA; dash-dotted blue: exact PIMC results using the full DLFC $G(q,\omega)$ taken from Ref.~\cite{dornheim_dynamic}; dashed black: \emph{static approximation} $G(q,\omega)\equiv G(q,0)$ using the static LFC from Ref.~\cite{dornheim_ML}; solid red [purple]: \emph{pair alignment} model, Eq.~(\ref{eq:dispersion}) using the effective KO potential [the effective PIMC potential, Eq.~(\ref{eq:Phi})]; solid yellow: classical one-component plasma at the same conditions.
Adapted from Dornheim \emph{et al.}~\cite{Dornheim_Nature_2022}.
}
\end{figure*}

\section{Applications\label{sec:applications}}
\subsection{Effective pair potential\label{sec:potential}}

Let us next utilize our new PIMC results for the effective force $F_\textnormal{eff}(r)$ to obtain the effective interaction potential between two electrons in the UEG medium. To this end, we numerically solve the simple one-dimensional integral Eq.~(\ref{eq:Phi}) and compare the results to the RPA potential and the KO potential with the PIMC-based SLFC in Fig.~\ref{fig:potential}. The left panel corresponds to $r_s=10$ and phenomenologically resembles the respective effective forces shown in Fig.~\ref{fig:compare} a). Interestingly, the Friedel oscillations in the RPA are less pronounced, whereas the overall systematic errors of RPA and the accurate linear-response prediction of the KO potential compared to the true PIMC results are similar to our results for $F_\textnormal{eff}(r)$. Similarly, the KO potential is quite accurate over the entire $r$-range, whereas the RPA exhibits more substantial deviations; the same trends can be found in Fig.~\ref{fig:potential}b) for the metallic density of $r_s=4$. Finally, the solid yellow curves show the classical potential of mean force, which we estimate by inserting our PIMC results for $g(r)$ into Eq.~(\ref{eq:phi_cl}). Evidently, the classical expression breaks down for small distances $r$, where quantum effects such as diffraction and exchange predominantly shape the physical behaviour of the system. Remarkably, though, $\phi_\textnormal{cl}(r)$ nicely captures the attractive minimum around $r=2r_s$ for $r_s=10$ and is substantially more accurate than the KO result.

\subsection{Spectrum of density fluctuations\label{sec:spectrum}}

Very recently, it has been shown that microscopic information about the effective electronic interaction gives straightforward access to the spectrum of density fluctuations $\omega(q)$~\cite{Dornheim_Nature_2022}. The latter is typically estimated in terms of the maximum of the dynamic structure factor $S(\mathbf{q},\omega)$, which can be expressed in the exact spectral representation as~\cite{quantum_theory}
\begin{eqnarray}\label{eq:spectral}
S(\mathbf{q},\omega) = \sum_{m,l} P_m \left\|{n}_{ml}(\mathbf{q}) \right\|^2 \delta(\omega - \omega_{lm})\ .
\end{eqnarray}
We note that $l$ and $m$ denote the eigenstates of the $N$-body Hamiltonian Eq.~(\ref{eq:Hamiltonian}), $\omega_{lm}=(E_l-E_m)/\hbar$ is their energy difference, and $n_{ml}$ is the corresponding transition element induced by the singe-particle density operator $\hat{n}(\mathbf{q})$. In other words, Eq.~(\ref{eq:spectral}) constitutes the sum over all possible transitions between the (time-independent) eigenstates of the system, with $P_m$ being the probability of the initial state. 

Naturally, a direct evaluation of Eq.~(\ref{eq:spectral}) constitutes a most formidable task and would, in principle, require the complete diagonalization of the many-body Hamiltonian. This is both unfeasible and not necessary; highly accurate results for $S(\mathbf{q},\omega)$ of the warm dense UEG have become available based on the analytic continuation of exact PIMC data for the imaginary-time density--density correlation function $F(\mathbf{q},\tau)$~\cite{dornheim_dynamic,dynamic_folgepaper,Hamann_PRB_2020,Dornheim_PRE_2020}. These results are based on the fully frequency-dependent DLFC $G(\mathbf{q},\omega)$, which allows one to resolve the interesting, nontrivial structure of $S(\mathbf{q},\omega)$ at intermediate wavenumbers, $q\sim2q_\textnormal{F}$, for moderate to strong coupling. For $r_s\gtrsim10$, this leads to the emergence of a \emph{roton feature} in $\omega(q)$, that is shown in Fig.~\ref{fig:dispersion} a). Specifically, the dash-dotted blue curve shows the exact PIMC results from Ref.~\cite{dornheim_dynamic} and the dotted green curve the estimation of $\omega(q)$ within the RPA. Clearly, the latter does not capture the pronounced down-shift in the dispersion curve around $q\sim2q_\textnormal{F}$, i.e., when the wave length $\lambda=2\pi/q$ becomes comparable to the average interparticle distance.

Based on the dramatic short-range nature of $\phi_\textnormal{eff}(r)$, Dornheim \emph{et al.}~\cite{Dornheim_Nature_2022} have proposed the \emph{electronic pair alignment} model. Within this framework, the \emph{roton feature} is caused by the excitation of electron pairs, which, in the presence of the electronic medium, interact via Eq.~(\ref{eq:Phi}). In particular, the probability to find two particles at a distance $r$ prior to the excitation, $P(r)=ng(r)$, plays the role of $P_m$ in Eq.~(\ref{eq:spectral}). The difference between the initial and final state is assumed to be restricted to this particle pair, which, after the excitation, are separated by a distance $\lambda$ along the direction of $\mathbf{q}$.  In the regime of the \emph{roton feature}, such pair excitations will effectively separate the particles, thereby reducing their interaction energy by the amount~\cite{Dornheim_Nature_2022}
\begin{eqnarray}\label{eq:shift}
\Delta{W}(q)=n\int\textnormal{d}\mathbf{r}g(r)\left[\phi_{\mathrm{eff}}(r)-\phi_{\mathrm{eff}}(r_{q})\right] \ .
\end{eqnarray} 
Let us briefly re-iterate the physical meaning of Eq.~(\ref{eq:shift}). The probability of finding a particle pair of distance $r$ in the unperturbed system is $P(r)=ng(r)$. Due to the excitation, the particles will be separated by the wavelength $\lambda$ after the excitation along the direction of the perturbation; the other coordinates remaining unchanged. The corresponding final separation is denoted as $r_q$. Clearly, this change in particle distance leads to a change in the effective potential energy $\Delta\phi_{\mathrm{eff}}(r,r_q)=\phi_{\mathrm{eff}}(r)-\phi_{\mathrm{eff}}(r_q)$. The \emph{average interaction energy change} $\Delta W(q)$ of a pair excitation of wavenumber $q$ is thus the average over all such transitions.

To estimate the red-shift in $\omega(q)$ compared to RPA, we have to estimate the corresponding exchange--correlation contribution
\begin{eqnarray}
\Delta W_\textnormal{xc}(q) &=& \Delta W(q) - \Delta W_\textnormal{RPA}(q)\ ,
\end{eqnarray}
where $W_\textnormal{RPA}(q)$ is obtained by inserting $g_\textnormal{RPA}(r)$ and $\phi_\textnormal{RPA}(r)$ into Eq.~(\ref{eq:shift}). Since the impact of exchange--correlation effects onto the kinetic energy is negligible at these conditions, we immediately get the ansatz
\begin{eqnarray}\label{eq:dispersion}
\omega(q) &=& \omega_\textnormal{RPA}(q) -  \Delta \omega_\textnormal{xc}(q)\ , \\
&=& \omega_\textnormal{RPA}(q) -  \alpha(q) \Delta W_\textnormal{xc}(q)\ . \nonumber
\end{eqnarray}
Here the screening function~\cite{kugler1} $\alpha(q) = \chi(q)/\chi_0(q)$ takes into account the fact that such pair excitations will become impossible for $\lambda\gg r_s$ due to the perfect screening in the UEG~\cite{kugler_bounds}.

In lieu of the true PIMC results for $\phi_\textnormal{eff}(r)$, Dornheim \emph{et al.}~\cite{Dornheim_Nature_2022} have used the linear-response expression due to Kukkonen and Overhauser, Eq.~(\ref{eq:KO}), and the results are shown as the solid red lines for $r_s=10$ [Fig.~\ref{fig:dispersion} a)] and $r_s=4$ [Fig.~\ref{fig:dispersion} b)]. It is evident that these curves capture the main part of the exchange--correlation induced down-shift compared to RPA in both cases. Interestingly, they closely follow the \emph{static approximation}~\cite{dornheim_dynamic} (dashed black), which has been obtained by setting the DLFC to its exact static limit, \emph{i.e.}, $G(\mathbf{q},\omega)\equiv G(\mathbf{q},0)$. This is expected, as both the \emph{static approximation} and Eq.~(\ref{eq:shift}) constitute a measure for an \emph{average energy shift}; see the original Ref.~\cite{Dornheim_Nature_2022} for a more detailed discussion of this point. 
Within the context of the present work, the most important result is illustrated as the purple curves in Fig.~\ref{fig:dispersion}, which has been obtained by evaluating Eq.~(\ref{eq:shift}) using as input our new PIMC results for the true effective potential, Eq.~(\ref{eq:Phi}). Overall, the results are in very good agreement to the red curves, which further substantiates the validity of the \emph{pair alignment} mechanism. In addition, we note that the true $\phi_\textnormal{eff}(r)$ leads to a small improvement towards the exact (dash-dotted blue) PIMC curves for both densities; for $r_s=10$, we even get a shallow \emph{roton feature} in the purple curve at the correct position that is not present in the results that have been obtained with the linear-response KO potential.

\section{Summary and Discussion\label{sec:summary}}

In this work, we have presented extensive new \emph{ab initio} PIMC results for the effective force $F_\textnormal{eff}(r)$ and the effective pair potential $\phi_\textnormal{eff}(r)$ between two electrons in the presence of the UEG. First and foremost, we have consistently observed a rapid decay of $F_\textnormal{eff}(r)$ towards large $r$, which is in stark contrast to the long-range tail of the bare Coulomb interaction. This has profound implications for computational quantum many-body theory, and explains the observed absence or the small magnitude of finite-size effects in wavenumber resolved properties, such as $S(q)$, that was observed in earlier studies of the UEG~\cite{Holzmann_PRB_2016,dornheim_prl,review,Chiesa_PRL_2006,Drummons_PRB_2008,dornheim_ML,Dornheim_JCP_2021}. Indeed, finite-size effects are negligible in both $F_\textnormal{eff}(r)$ and $\phi_\textnormal{eff}(r)$ for as few as $N=14$ electrons at WDM conditions. 

From a physical perspective, our study has provided the numerical proof for an effective attraction between two electrons around $r\sim2r_s$ for $r_s\gtrsim5$ and $\theta\lesssim2$. We re-iterate our earlier point that this does \emph{not} arise due to the mediation through phonons or other ionic effects, and that the rigid, non polarizeable background in the UEG model cannot be the origin of this effect. Instead, it is a direct consequence of the incipient localization of the electrons upon entering the electron liquid regime. 

The comparison of our new PIMC simulation data to analytical theories has revealed that the widely used random phase approximation does not provide an adequate description of the microscopic interaction between a pair of UEG electrons at metallic densities. In stark contrast, the linear-response expression by Kukkonen and Overhauser~\cite{Kukkonen_PRB_1979} allows us to consistently incorporate electronic exchange--correlation effects based on previous PIMC data for the static local field correction. This leads to a dramatic improvement over the RPA, and accurately captures the main trends of the exact PIMC results. At the same time, the KO results overestimate the depth of the attractive minimum in both $F_\textnormal{eff}(r)$ and $\phi_\textnormal{eff}(r)$. This is a direct consequence of the correlational origin of the effective attraction, which, therefore, requires a more advanced description that incorporates nonlinear terms~\cite{Dornheim_JPSJ_2021}. In addition, we have found substantial deviations between the true fermionic PIMC estimator for the effective force and the naive histogram estimator for $r<r_s$, where the former approaches the bare Coulomb repulsion, whereas the latter is comparably reduced in magnitude. These deviations have been explained by the fermionic degeneracy pressure, which is consistently removed from the true force, but biases the simple histogram implementation.
%To understand the physical origin of this effect, we have carried out PIMC simulations of \emph{distinguishable Boltzmann particles}, which include all nonlinear and quantum diffraction effects, but omit the fermionic anti-symmetry under the exchange of particle coordinates. Evidently, the Boltzmann simulations reproduce the analytical curves in the limit of small $r$, which means that the reduction in $F_\textnormal{eff}(r<r_s)$ in the true PIMC results is due to fermionic exchange effects. The additional analysis of the spin-resolved components of $F_\textnormal{eff}(r)$ has further substantiated this interpretation.

Finally, we have used our new PIMC results for the effective electronic pair interaction $\phi_\textnormal{eff}(r)$ to estimate the spectrum of density fluctuation $\omega(q)$ within the recently introduced \emph{electronic pair alignment} model. Our results further validate previous data that were based on the linear-response KO potential, and constitute a small improvement over the latter. We, thus, safely conclude that the studied effective interaction gives one direct microscopic insight into the origin of the \emph{roton feature} in the warm dense UEG.

We are convinced that our investigation will open up many avenues for important and impactful future research, and briefly sketch a number of promising possibilities: i) our PIMC results give one direct insight into the microscopic impact of \emph{nonlinear effects} onto the effective interaction between a pair of electrons. This can be used as the basis for the further improvement of dielectric theories~\cite{tanaka_hnc,Tolias_JCP_2021,castello2021classical,arora} and guide the development of explicitly nonlinear potentials~\cite{nonlinear1,Gravel}; ii) the possible role of the attractive minimum in $F_\textnormal{eff}(r)$ as a \emph{pairing mechanism} in the UEG, and its connection to the potential emergence of superconductivity in the electron liquid regime~\cite{Takada_PRB_1993} deserve closer exploration; iii) our analysis has revealed the remarkably fast decay of fermionic exchange effects for $r\sim r_s$, which opens up the intriguing possibility for a further decomposition: use of fermionic PIMC for very small $N$ to study short range exchange--correlation effects at $r\lesssim r_s$, use of Boltzmann PIMC to study diffraction at $r_s \lesssim r \lesssim 3r_s$, and use of dielectric theories  for $r\gg r_s$ to get a highly accurate description across the full domain of length scales. This might substantially increase the scope of feasible density-temperature combinations~\cite{Yilmaz_JCP_2020}, and give us for the first time access to the regime of metallic densities at $\theta\sim0.1,\dots,0.5$; iv) it is straightforward to estimate both $F_\textnormal{eff}(r)$ and $\phi_\textnormal{eff}(r)$ for a number of other systems that can be simulated with the PIMC approach. This can give new insights into important phenomena like superfluidity~\cite{cep,Ferre_Boronat_PRB_2016,griffin1993excitations} and supersolid behaviour~\cite{RevModPhys.84.759,Saccani_Supersolid_PRL_2012,PhysRevLett.122.130405,Norcia2021}. In addition, we suggest to use the \emph{pair alignment} model~\cite{Dornheim_Nature_2022} to explain the spectrum of density fluctuations in other systems, with the \emph{original roton feature} in ultracold $^4$He~\cite{Ferre_Boronat_PRB_2016,Boninsegni1996,griffin1993excitations} and $^3$He~\cite{Godfrin2012,Dornheim_SciRep_2022,PhysRevLett.37.842,Nava_PRB_2013}
being a promising candidate; v) finally, we will extend the present study beyond the UEG model, starting with hydrogen in the WDM regime~\cite{Militzer_PRE_2001,Bohme_PRL_2022}. This will allow us to investigate the effective interaction between all individual components, and provide us with unprecedented insights into the behaviour of WDM at the nanoscale.

\renewcommand{\theequation}{A\arabic{equation}}
\setcounter{equation}{0}
\appendix
\section*{Appendix: Classical relation between the Kukkonen-Overhauser potential and the potential of mean force}

In what follows, we shall derive an exact classical relation between the Kukkonen-Overhauser potential and the potential of mean force. This classical connection is not restricted to Coulomb interactions but it is valid for arbitrary pair interaction potentials (even anisotropic). It can be exploited in order to systematically improve the effective interaction predictions of linear response theory without significant effort.

The general static form of the effective KO potential in the unpolarized case remains intact in the classical limit. Irrespective of the pair interaction potential, it reads as
\begin{equation*}
\beta\Phi^{\mathrm{KO}}_{\mathrm{eff}}(\boldsymbol{q})=\beta{\Phi}(\boldsymbol{q})+\frac{1}{n}\left\{n\beta{\Phi}(\boldsymbol{q})[1-G(\boldsymbol{q})]\right\}^2\frac{\chi(\boldsymbol{q})}{n\beta}\,.
\end{equation*}
The classical version of the fluctuation-dissipation theorem, the zeroth frequency moment rule and the Kramers-Kronig relation lead to the well-known expression $S(\boldsymbol{q})=-\chi(\boldsymbol{q})/(n\beta)$. Together with the static limit of the ideal Vlasov density response $\chi_0(\boldsymbol{q})=-n\beta$ and static limit of the general relation for the density response, Eq.\ref{eq:chi_tot}, the above expression yields $n\beta{\Phi}(\boldsymbol{q})[1-G(\boldsymbol{q})]=[1/S(\boldsymbol{q})]-1$. Thus,
\begin{equation}
\beta\Phi^{\mathrm{KO}}_{\mathrm{eff}}(\boldsymbol{q})=\beta{\Phi}(\boldsymbol{q})-\frac{1}{n}\frac{[1-S(\boldsymbol{q})]^2}{nS(\boldsymbol{q})}\,.
\end{equation}
In order to further simplify the above, the Fourier space connection between the static structure factor and total correlation function $S(\boldsymbol{q})=1+nH(\boldsymbol{q})$ is employed and the Fourier transformed exact Ornstein-Zernike equation $H(\boldsymbol{q})=C(\boldsymbol{q})+nH(\boldsymbol{q})C(\boldsymbol{q})$ is utilized, where $C(\boldsymbol{q})$ is the Fourier transformed direct correlation function $c(\boldsymbol{r})$. The resulting expression can be trivially inverted back to real space and reads as
\begin{equation*}
\beta\phi^{\mathrm{KO}}_{\mathrm{eff}}(\boldsymbol{r})=\beta\phi(\boldsymbol{r})+c(\boldsymbol{r})-h(\boldsymbol{r})\,.
\end{equation*}
Recall that the exact non-linear closure equation of the integral equation theory of classical liquids is given by $g(\boldsymbol{r})=\exp{[-\beta\phi(\boldsymbol{r})-c(\boldsymbol{r})+h(\boldsymbol{r})+b(\boldsymbol{r})]}$\,\cite{liquid_state_Lee,liquid_state_Santos}, with $b(\boldsymbol{r})$ the bridge function. Application of the natural logarithm, disposal of the common $\beta\phi(\boldsymbol{r})+c(\boldsymbol{r})-h(\boldsymbol{r})$ expression and substitution for the potential of mean force, ultimately yield
\begin{equation}\label{eq:appendix}
\beta\phi^{\mathrm{KO}}_{\mathrm{eff}}(\boldsymbol{r})=\beta\phi_{\mathrm{cl}}(\boldsymbol{r})+b(\boldsymbol{r})\,.
\end{equation}
Note that, thus far, only exact classical expressions have been employed, which implies that $\beta\phi_{\mathrm{cl}}(\boldsymbol{r})$ is the exact effective interaction. The above expression demonstrates that the effective KO potential is only exact within the hypernetted chain approximation (HNC), which assumes that the bridge function is zero, $b(\boldsymbol{r})=0$. Hence, in the classical limit, it can be deduced that the use of linear response theory is exactly equivalent to the approximation that the contribution of highly connected irreducible diagrams (which constitute the bridge function) to the pair correlation function is negligible. Treatment of the general quantum case with diagrammatic perturbation theory has revealed that the KO derivation implicitly contains two additional approximations~\cite{Vignale_Singwi_PRB_1985,quantum_theory}: the substitution of the internal electron-hole propagator by non-interacting electron-hole propagators and the replacement of the irreducible electron-hole interaction by some kind of an average.

In the dilute strongly coupled uniform electron fluid regime, $r_{\mathrm{s}}\gtrsim20$,
Eq.~(\ref{eq:appendix}) can be employed to improve the effective potential predictions of the KO expression. At lower densities, fermionic exchange effects are far less pronounced and the same should apply for the consequences of the implicit approximations of the KO derivation. Simultaneously, irreducible diagram effects become significant but they can be treated classically, as deduced from the success of the integral equation theory based dielectric scheme at strong coupling~\cite{Tolias_JCP_2021,castello2021classical}. Given the availability of a closed-form expression for the bridge function of the classical one component plasma at arbitrary coupling~\cite{castello2021classical,LuccoCastello2022}, $\beta\phi_{\mathrm{eff}}(\boldsymbol{r})=\beta\phi^{\mathrm{KO}}_{\mathrm{eff}}(\boldsymbol{r})-b(\boldsymbol{r})$ should constitute an improvement over the original KO expression. This approximation will be explored in a future work that will present PIMC results beyond the WDM regime.

\section*{Acknowledgments}
This work was partially supported by the Center for Advanced Systems Understanding (CASUS) which is financed by Germany’s Federal Ministry of Education and Research (BMBF) and by the Saxon state government out of the State budget approved by the Saxon State Parliament.
The PIMC calculations were carried out at the Norddeutscher Verbund f\"ur Hoch- und H\"ochstleistungsrechnen (HLRN) under grant shp00026, and on a Bull Cluster at the Center for Information Services and High Performance Computing (ZIH) at Technische Universit\"at Dresden.

%%%%%%%%%%%%%%%%%%%%%%%%%%%%%%%%%%%%%%%%%%%%%%%%%%%%%%%%%%%%%%%%%%%%%%%%%%%%%%%%
% literature
%%%%%%%%%%%%%%%%%%%%%%%%%%%%%%%%%%%%%%%%%%%%%%%%%%%%%%%%%%%%%%%%%%%%%%%%%%%%%%%%
\bibliography{bibliography}

\begin{thebibliography}{157}
\bibitem [{rep()}]{repo}%
  \BibitemOpen
  \href@noop {} {}\bibinfo {note} {A link to a repository containing all PIMC
  raw data will be made available upon publication.}\BibitemShut {Stop}%
\end{thebibliography}
\end{document}